\documentclass[a4paper,11pt]{article}
\usepackage{jcappub}
\usepackage{graphicx}
\usepackage{amsmath}
\usepackage{amsmath}
\usepackage{xspace}
\usepackage{subfig}
\usepackage{float}
\usepackage{color}
\usepackage[utf8]{inputenc}
\usepackage{commath}
\usepackage{slashed}
\usepackage{fancyvrb}
\usepackage[utf8]{inputenc}
\usepackage[english]{babel}
\usepackage{amsfonts}
\usepackage{amssymb}
\usepackage{orcidlink}
\usepackage{color}
\usepackage{tabularray}
\UseTblrLibrary{booktabs}
\title{Revisiting the limits on dark matter annihilation cross-section and decay lifetime in light of electron and positron fluxes}
\author[a,b]{Nagisa Hiroshima,}
\emailAdd{hiroshima-nagisa-hd@ynu.ac.jp}

\author[c,d,e,f]{Kazunori Kohri,}
\emailAdd{kazunori.kohri@gmail.com}

\author[g]{Partha Kumar Paul\orcidlink{0000-0002-9107-5635},}
\emailAdd{ph22resch11012@iith.ac.in}

\author[g]{and Narendra Sahu\orcidlink{0000-0002-9675-0484}}
\emailAdd{nsahu@phy.iith.ac.in}

\affiliation[a]{Department of Physics, Faculty of Engineering Science, Yokohama National University, Yokohama 240–8501, Japan}

\affiliation[b]{RIKEN Center for Interdisciplinary Theoretical and Mathematical Sciences(iTHEMS), RIKEN,
Wako 351-0198, Japan}

\affiliation[c]{Division of Science, National Astronomical Observatory of Japan (NAOJ),
and SOKENDAI, 2-21-1 Osawa, Mitaka, Tokyo 181-8588, Japan}

\affiliation[d]{Department of Astronomy, The University of Tokyo, Bunkyo-ku, Hongo, Tokyo 113-0033, Japan}

\affiliation[e]{Theory Center, IPNS, KEK, 1-1 Oho, Tsukuba, Ibaraki 305-0801, Japan}

\affiliation[f]{Kavli IPMU (WPI), UTIAS, The University of Tokyo, Kashiwa, Chiba 277-8583, Japan}

\affiliation[g]{Department of Physics, Indian Institute of Technology Hyderabad, Kandi, Telangana-502285, India.}

\abstract{We revisit the upper bound on the annihilation cross-section, $\langle\sigma v\rangle$ of a stable dark matter (DM) of mass $500-10^{14}$ GeV by considering five different channels: $W^+W^-$, $b\bar{b}$, $\mu^+\mu^-$, $\tau^+\tau^-$, and $e^+e^-$. We use the observed electron and positron fluxes from CALET, DAMPE, HESS, positron flux from AMS-02,  and gamma-ray flux from HAWC, GRAPES-3, CASA-MIA to constrain the annihilation cross-section. We also consider unstable DM of mass $10^3-10^{16}$~GeV decaying to $W^+W^-$, $b\bar{b}$, $\mu^+\mu^-$, $\tau^+\tau^-$, and $e^+e^-$ and derive the corresponding lower bound on the DM lifetime, $\tau_{\rm DM}$. We find that the latest AMS-02 data provide the most stringent constraints on $\langle\sigma v\rangle$ for DM masses below 2 TeV, while HESS yields the strongest limits for $M_{\rm DM}\gtrsim2$ TeV. The HESS gives a much more stringent limit on the DM lifetime, excluding $\tau_{\rm DM\rightarrow\mu^+\mu^-}\lesssim\mathcal{O}(10^{30})$ s for a 10 TeV mass of DM. The limits on $\langle\sigma v\rangle$ derived from the $e^+e^-$ flux are competitive with those from $\gamma$-ray and neutrino observations for DM masses in the range $10^5$--$10^{11}$ GeV, and become the most stringent beyond this range. For decaying DM, the $e^+e^-$ flux provides the strongest constraints on the DM lifetime over the mass range $10^3$--$10^9$ GeV.}

\arxivnumber{2510.11700}

\makeatletter
\gdef\@fpheader{To appear in JCAP\hfill
	\begin{minipage}[t]{5cm}
		\raggedleft\upshape
		KEK-TH-2768\\
		KEK-Cosmo-0395\\
		RIKEN-iTHEMS-Report-25
\end{minipage}}
\makeatother

\begin{document}
\maketitle
\flushbottom
\section{Introduction}

Various astrophysical and cosmological observations, such as galaxy rotation curves, gravitational lensing, bullet clusters, the cosmic microwave background, etc, establish the existence of an invisible component of the Universe, called dark matter (DM) via its gravitational interaction \cite{Zwicky:1933gu,Rubin:1970zza,Clowe:2006eq,Planck:2018vyg}. The existence of large scale structures of the Universe imply that the DM is either stable or the lifetime is longer than the age of the Universe if it is unstable. Such a particle is absent within the standard model (SM) of particle physics. Assuming that the DM is a fundamental particle, its mass, spin, and its interactions with the SM particles apart from gravity are yet a mystery. Determining these properties is one of the most important open challenges at the interface of particle physics and cosmology (See Ref.~\cite{Bertone:2004pz,Bertone:2018krk,Profumo:2019ujg} for examples). 

Among the various search strategies, indirect detection of DM plays a crucial role. If DM particles annihilate or decay to produce SM particles, those particles could leave interesting signatures in either in the early or current Universe. Electrons ($e^-$) and positrons ($e^+$) are especially interesting messengers in this context (see for instance~\cite{Bergstrom:2009fa,Ibarra:2013cra,Slatyer:2015jla}). They carry valuable information about nearby sources, including a possible DM origin. Over the past decade, a series of precision measurements of the cosmic-ray $e^++e^-$ spectrum by different experiments such as CALET~\cite{CALET:2023emo}, DAMPE~\cite{Alemanno:2022eeb}, H.E.S.S~\cite{HESS:2024etj}, and the $e^+$ spectrum by AMS-02~\cite{AMS:2021nhj} have opened up the possibility of probing DM models with greater sensitivity. In addition, the gamma-ray data reported by HAWC~\cite{HAWC:2022uka}, GRAPES-3~\cite{grapes:2009}, and CASA-MIA \cite{CASA-MIA:1997tns} also provide valuable information on $e^++e^-$ pairs since they cannot distinguish gamma-ray from $e^++e^-$ pairs \footnote{We provide the details of the flux data in Appendix \ref{app:data}.}. These data can be utilized to place upper bounds on the annihilation cross section and lower bounds on the lifetime of dark matter in the $e^+e^-$ channel. These datasets span a wide range of energies and provide complementary coverage. 

Local astrophysical sources are also possible to produce electrons and positrons, such as pulsars \cite{Hooper:2008kg,Fang:2018qco} and supernova remnants~\cite{Kohri:2015mga,Blasi:2009bd,Mertsch:2014poa,Mertsch:2020ldv,DiMauro:2020cbn} within the Milky Way galaxy. Previous works have investigated their contributions in detail and assessed that such contributions can be neglected for discussions on DM properties. Many efforts have been devoted to constraining the DM annihilation cross-section and lifetime using data from $e^+$, $e^-$, gamma rays, antiprotons, and neutrinos; see \textit{e.g.} \cite{Feng:2013zca,Boudaud:2016mos,MAGIC:2022acl,HESS:2022ygk,Song:2024vdc,Cheng:2023chi,Calore:2022stf,Kohri:2025bsn,Boehm:2025qro,Nguyen:2024kwy,Nguyen:2025tkl,Dutta:2022wdi,Dutta:2022wuc,Dubey:2025ouh,Maity:2021umk,LHAASO:2024upb,Chianese:2026cfz} 

In this work, we study two scenarios: DM annihilation and DM decay into various SM final states that subsequently produce electrons and positrons, focusing on the relatively heavy regime of $M_{\rm DM}\gtrsim500$~GeV, where constraints from underground or collider experiments are relatively weak~\cite{GAMBIT:2017snp}. For each case, we derive 95\% C.L. limits on the annihilation cross-section, $\langle\sigma v\rangle$, and the lifetime, $\tau_{\rm DM}$, as functions of the DM mass in the range $500~{\rm GeV}\leq M_{\rm DM}\leq10^{16}~{\rm GeV}$. Previous works constrain parts of the region of $M_{\rm DM}\lesssim10^{11}$GeV using gamma-ray or neutrino observations~\cite{Song:2024vdc, Boehm:2025qro, Cohen:2016uyg, Kohri:2025bsn}, however, at the regime of $M_{\rm DM}\gtrsim10^{11}$GeV, only a limited number of works (e.g~\cite{Das:2023wtk}) or prospects ~\cite{Chianese:2021htv} are available. In this work, for the first time, we derive limits on DM up to $M_{\rm DM}=10^{16}$GeV using electron and positron observations.

The paper is organized as follows. In section \ref{sec:epflux}, we discuss the electron and positron flux generation from the annihilation and decay of a DM in general. The methodology and results are discussed in section \ref{sec:result}. We finally conclude in section \ref{sec:conlc}.

\section{Electron and Positron Excesses from DM Annihilation and decay}\label{sec:epflux}

\subsection{Production and propagation of positron and electron}
The annihilation or decay of DM to various SM states can give rise to electrons and positrons in the Galactic medium. Once the electrons and positrons are produced, they travel in the Galaxy under the influence of the Galactic magnetic field, which is assumed to be of the order of a few micro-gauss \cite{Kronberg:1993vk,Zhang:2008rs}. As a result, their motion can be 
thought of as a random walk. Thus, a fraction of the electron and positron flux comes out and reaches our solar 
system. The electron or positron flux in the vicinity of our solar system can be obtained by solving the diffusion 
equation~\cite{Moskalenko:1997gh,Delahaye:2007fr}
\begin{eqnarray}
\frac{\partial }{\partial t} f_{e^\pm}(E,\vec{r},t)&=&
K(E) \nabla^2 f_{e^\pm}(E,\vec{r},t)+ 
\frac{\partial}{\partial E}[b(E) f_{e^\pm}(E,\vec{r},t)]+  Q(E,\vec{r})
\label{diffusion}  
\end{eqnarray}
where $f_{e^\pm}(E,\vec{r})$ is the number density of positrons (electron) per unit energy,
$E$ is the energy of positron (electron), $K(E)$ is the diffusion coefficient, $b(E)$ is the energy-loss
rate\footnote{For a more detailed method,
see \cite{John:2023ulx}.} and $Q(E,\vec{r})$ is the positron (electron) source term, which is given by:
\begin{equation}
Q(E,\vec{r})=\left\{
		\begin{array}{l}
			\frac{1}{2}n_{\rm DM}^2(\vec{r})f_{\rm inj}^{{\rm ann}}\\
			n_{\rm DM}(\vec{r})f_{\rm inj}^{{\rm dec}}\\
		\end{array}
		\right.
       \label{eq:source}
\end{equation}
where the injection spectrum $f_{\rm inj}^{\rm ann},f_{\rm inj}^{\rm dec}$ can be given by
\begin{eqnarray}
 f_{\rm inj}^{{\rm ann}}= \langle \sigma v\rangle\frac{d N_{e^\pm}}{d E};~
  f_{\rm inj}^{{\rm dec}}=\frac{1}{\tau_{\rm DM}}\frac{d N_{e^\pm}}{d E}.
\label{injection}   
\end{eqnarray}
The factor of 1/2 in Eq.(\ref{eq:source}) arises from the assumption that DM is self-conjugate, \textit{i.e.}, identical to its own antiparticle.
In the above equation the energy spectrum $d N_{e^\pm}/d E$ represents the
number of positrons or electrons with energy $E$, which are produced from the annihilation or decay of DM, and 
$\langle \sigma v \rangle$ is the thermal averaged annihilation cross-section, and $\tau_{\rm DM}$ is the decay lifetime of the DM, which characterizes the model.

In the low velocity limit, there is the so-called unitarity bound on the $s$-wave annihilation cross-section which is given as
\cite{Griest:1989wd,Hui:2001wy,Beacom:2006tt}
\begin{eqnarray}
    \langle\sigma v\rangle\leq1.5\times10^{-13}{\rm cm^{3}/s}\left( \frac{1~{\rm GeV}}{M_{\rm DM}} \right)^2\left( \frac{300~{\rm km/s}}{v} \right).
\end{eqnarray}
The unitarity bound is typically derived under the assumption that the DM mass and the mediator mass are of the same order, as realized in various 4-dimensional supersymmetric theories scenarios such as the minimal supersymmetric Standard Model (MSSM), constrained MSSM, or minimal supergravity(mSUGRA)~\cite{Martin:1997ns}. In such cases, the bound provides a meaningful upper limit on the viable DM mass. However, if the mediator is much lighter than the DM, the standard unitarity argument no longer applies. In this regime, the annihilation cross-section can be enhanced, thereby relaxing the unitarity constraint \cite{Arina:2010wv}\footnote{In the presence of a light mediator, the latter gives a long-range attractive potential which leads to Sommerfeld enhancement in the annihilation cross-section, especially when the relative velocity between the DM particles is small. Since this enhancement arises from a highly non-perturbative process, the usual unitarity limit does not apply.} Since the light mediator is not itself a DM candidate, it may be unstable, underproduced in the early Universe, or appear only as a virtual state. The unitarity bound does not impose a fundamental limit on the DM mass in these scenarios. Another way the unitarity limit can be evaded is if one considers the DM to be an extended object with a finite radius $R$ instead of a point-like particle ($R=0$). In such a scenario, the unitarity limit on the annihilation cross-section is modified to $\langle\sigma v\rangle\leq4\pi(1+M_{\rm DM}Rv)^2/M_{\rm DM}^2v$ \cite{Griest:1989wd,Frumkin:2022ror,Boehm:2025qro}.

We now discuss the solution of Eq.(\ref{diffusion}). We assume that the positrons (electrons) are in a steady state, i.e., $\partial f_{e^\pm}/\partial t=0$. Then from 
Eq. (\ref{diffusion}), the positron (electron) flux in the vicinity of our solar system can be obtained in a 
semi-analytical form~\cite{Delahaye:2007fr,Hisano:2005ec,Cirelli:2008id}
\begin{equation}
\Phi_{e^\pm} (E,\vec{r}_{\odot}) =\frac{c}{4\pi b(E)}\left\{
		\begin{array}{l}
			\frac{1}{2}(n_{\rm DM})_{\odot}^2 
\langle \sigma v \rangle\int_E^{M_{\rm DM}} dE_s \frac{dN_{e^\pm}}{dE_s}.\mathcal{I} (\lambda_D(E,E_s))\\
			(n_{\rm DM})_{\odot} 
\frac{1}{\tau_{\rm DM}}\int_E^{M_{\rm DM}/2} dE_s \frac{dN_{e^\pm}}{dE_s}.\mathcal{I} (\lambda_D(E,E_s))\\
		\end{array}
		\right.
        \label{positron_flux}
\end{equation}
where $(n_{\rm DM})_\odot=\rho_\odot/M_{\rm DM}$, $\lambda_D(E,E_s)$ is the diffusion length from energy $E'$ to energy $E$ and $\mathcal{I}(\lambda_D(E,E_s)$
is the halo function, which is independent of particle physics. The Halo function is given as \cite{Genolini:2021doh}
\begin{eqnarray}
    \mathcal{I}(\lambda_{D})=\int_{\mathcal{V}_L^{\odot}}d^3\vec{x}_{s}\frac{e^{-\frac{|\vec{x}_{\odot}-\vec{x}_{s}|^2}{2\lambda_D^2}}}{(2\pi^2\lambda_D^2)^{3/2}}\left( \frac{\rho(\vec{x}_{s})}{\rho_\odot} \right)^m,
\end{eqnarray}
where $m=2$ for annihilation and $m=1$ for decay.
$\lambda_D$ is calculated using
\begin{eqnarray}
\lambda^2_D(E,E_s)=2\int_E^{E_s}dE'\frac{K(E')}{b(E')},
\end{eqnarray}
with $b(E)=\frac{E^2}{E_*\tau_E}$ where $E_*=1$ GeV is a reference energy and $\tau_E=10^{16}$ s. The diffusion coefficient is given by
\begin{eqnarray}
    K(E)=K_0\left[ 1+\left( \frac{R_1}{E} \right)^{\frac{-\delta_1+\delta}{s_1}} \right]^{s_1}E^\delta\left[ 1+\left( \frac{E}{R_h} \right)^{\frac{\Delta_h}{s_h}} \right]^{-s_h},\label{eq:diffcoff}
\end{eqnarray}
\begin{table}[h]
\begin{center}
		\begin{tblr}{
					colspec={|l|l|},
					 row{1}={font=\bfseries}
				}
				\toprule Parameters & SLIM\\
				\toprule
				$s_1$ & 0.05\\
                \hline
				$\Delta_h$& 0.19\\
                \hline
				$R_h$ [GV]& 237 \\\hline
				$s_h$& 0.04 \\
				\bottomrule
		\end{tblr}
		\caption{Parameter values for the SLIM configuration \cite{Genolini:2021doh}.}
		\label{tab:tab0}
\end{center}
\end{table}
where $R_1$ is the low-rigidity break, $\delta$ is the diffusion index, $\delta_1$ low-rigidity break index. The values of the parameters in Eq. \ref{eq:diffcoff} are listed in Table \ref{tab:tab0} and \ref{tab:tab00} for the SLIM configuration \cite{Genolini:2021doh}.
\begin{table}[h]
\begin{center}
		\begin{tblr}{
					colspec={|l|l|l|l|l|l|},
					 column{1}={font=\bfseries}
				}
				\toprule SLIM & $L$ [kpc]& $\delta$& $\log_{10}K_0$ [$\rm kpc^2Myr^{-1}$]& $R_1$ [GV]& $\delta_1$ \\
				\toprule
				{\tt MIN} & 2.56&0.509&-1.71&4.21 &-1.45\\
                \hline
				{\tt MED} & 4.67&0.499&-1.44&4.48 &-1.11\\
                \hline
				{\tt MAX} & 8.40&0.490&-1.18&4.74 &-0.776\\
				\bottomrule
		\end{tblr}
		\caption{Propagation parameters corresponding to the {\tt MIN}, {\tt MED}, and {\tt MAX} benchmarks for SLIM configuration \cite{Genolini:2021doh}.}
		\label{tab:tab00}
\end{center}
\end{table}
The DM density profile can be parametrized as
\begin{eqnarray}
    \rho(r)=\rho_\odot\left( \frac{r_\odot}{r} \right)^\gamma \left[ \frac{1+\left(\frac{r_\odot}{r_p}\right)^\alpha}{1+\left(\frac{r}{r_p}\right)^\alpha} \right]^{\frac{\beta-\gamma}{\alpha}},
\end{eqnarray}
with the parameter values corresponding to the NFW profile are given in Table \ref{tab:tab000}.
In the cylindrical co-ordinate Earth is located at $\vec{x}=(R_\odot,0,0)$ and the positron(electron) source is at $\vec{x}_s=(R_s,\phi_s,Z_s)$. Then the Halo function can be expressed as
\begin{eqnarray}
    \mathcal{I}=\int_0^{R_{\rm gal}}dR_s\int_{-L}^{+L}dZ_s\int_0^{2\pi}d\phi_s\frac{R_s}{(2\pi\lambda_D^2)^{3/2}}e^{-\frac{R_s^2+R_\odot^2-2R_sR_\odot\cos\phi_s+Z_s^2}{2\lambda_D^2}} \left( \frac{\rho(\sqrt{R_s^2+Z_s^2})}{\rho_\odot} \right)^m.
\end{eqnarray}
The angular integration results into
\begin{eqnarray}
    \mathcal{I}=2\pi\int_0^{R_{\rm gal}}dR_s\int_{-L}^{+L}dZ_s\frac{R_s}{(2\pi\lambda_D^2)^{3/2}}e^{-\frac{R_s^2+R_\odot^2+Z_s^2}{2\lambda_D^2}} \mathcal{I}_0\left( \frac{R_sR_\odot}{\lambda_D^2} \right) \left( \frac{\rho(\sqrt{R_s^2+Z_s^2})}{\rho_\odot} \right)^m,
\end{eqnarray}
where $\mathcal{I}_0\left( \frac{R_sR_\odot}{\lambda_D^2} \right)$ is the zeroth order modified Bessel function of first kind. Now the Halo function can be written as
\begin{eqnarray}
    \mathcal{I}&=&2\pi\int_0^{R_{\rm gal}}dR_s\frac{R_s}{(2\pi\lambda_D^2)}e^{-\frac{R_s^2+R_\odot^2}{2\lambda_D^2}} \mathcal{I}_0\left( \frac{R_sR_\odot}{\lambda_D^2} \right)\nonumber\\&&\times\int_{-L}^{+L}dZ_s \frac{1}{\sqrt{2\pi}\lambda_D} e^{-\frac{Z_s^2}{2\lambda_D^2}} \left( \frac{\rho(\sqrt{R_s^2+Z_s^2})}{\rho_\odot} \right)^m.
\end{eqnarray}
Since diffusion occurs within a magnetic halo of finite half-height $L$, particles reaching $Z_s=\pm L$ escape the Galaxy. The infinite-space expression therefore overestimates the flux when $\lambda_D\gtrsim L$. We account for this by imposing the boundary condition at $Z_s=\pm L$ via the method of images, which effectively pinches the halo function and suppresses it at large diffusion lengths. The Halo function is then given as
\begin{eqnarray}
    \mathcal{I}&=&2\pi\int_0^{R_{\rm gal}}dR_s\frac{R_s}{(2\pi\lambda_D^2)}e^{-\frac{R_s^2+R_\odot^2}{2\lambda_D^2}} \mathcal{I}_0\left( \frac{R_sR_\odot}{\lambda_D^2} \right)\nonumber\\&&\times\int_{-L}^{+L}dZ_s \left[\frac{1}{\sqrt{2\pi}\lambda_D} \sum_{n=-\infty}^{+\infty}(-1)^n e^{-\frac{\left(2nL+(-1)^n Z_s\right)^2}{2\lambda_D^2}} \right]\left( \frac{\rho(\sqrt{R_s^2+Z_s^2})}{\rho_\odot} \right)^m\label{eq:Ifinal}
\end{eqnarray}
\begin{table}[h]
\begin{center}
		\begin{tblr}{
					colspec={|l|l|l|l|l|},
					 column{1}={font=\bfseries}
				}
				\toprule Halo profile & $\alpha$& $\beta$& $\gamma$& $r_p$ [kpc]\\
				\toprule
				NFW & 1&3&1&18.6\\
				\bottomrule
		\end{tblr}
		\caption{DM Halo profile parameters.}
		\label{tab:tab000}
\end{center}
\end{table}
Thus, we ensure that for $Z_S>\pm L$, $\mathcal{I}=0$. We fix the Earth distance from the galactic center, $R_\odot=8.21$ kpc, local DM density $\rho_\odot=0.383{~\rm GeVcm^{-3}}$, and $R_{\rm gal}=20$ kpc.

By inserting the energy spectrum in Eq.(\ref{positron_flux}), one can calculate the total positron (electron) flux using 
the above expressions. For our analysis, we adopt the MIN propagation model together with an NFW profile for the Galactic dark matter halo. The corresponding parameters are listed in Table \ref{tab:tab00} and \ref{tab:tab000}. Our assumption of the NFW-{\tt MIN} model gives conservative limits regarding the fact that the distribution of DM in the central region of the Galaxy is subject to substantial astrophysical uncertainties \footnote{To estimate the uncertainty associated with the Galactic propagation model, we compare the results obtained using the MIN, and MAX benchmark configurations. We find that the inferred annihilation cross section (DM lifetime) varies by less than $\sim10$\% across these propagation scenarios in the mass range considered. Therefore, the impact of propagation uncertainties on our conclusions is negligible. See Appendix \ref{app:flux} for more details.}. The {\tt MIN} propagation model, characterized by a smaller diffusive halo height $L$ and diffusion coefficient $K_0$ compared to {\tt MED} and {\tt MAX}, yields the minimum predicted $e^\pm$ flux at Earth for a fixed annihilation cross-section. This approach ensures that our derived constraints remain robust against halo-profile uncertainties and are not overly optimistic for potential enhancements near the Galactic Center.

We consider very heavy DM in the mass range $500~{\rm GeV}\leq M_{\rm DM}\leq10^{16}~{\rm GeV}$ under the assumption that each DM particle decays into a pair of SM particles or DM particles annihilate to SM particles. The annihilation or decay channels considered are $W^+W^-$, $b\bar{b}$, $\mu^+\mu^-$, $\tau^+\tau^-$, and $e^+e^-$. The energy spectra are calculated using {\tt HDMSpectra}~\cite{Bauer:2020jay}, which incorporates the complete electroweak corrections, which is an essential ingredient in the study of heavy DM.

\subsection{Background fluxes of electron and positron}

The electron or positrons in our Galaxy are not only produced by DM annihilation or decay but also by the
scattering of cosmic-ray protons with the interstellar medium~\cite{Moskalenko:1997gh}.
Thus, the electron or positrons produced from the later sources can act as the background for the electron or positrons produced
from the annihilation or decay of DM. The background fluxes~\cite{Moskalenko:1997gh}
of primary and secondary electrons and secondary positrons can be parameterized
as~\cite{Baltz:1998xv}
\begin{eqnarray}
\Phi_{\rm prim,\; e^-}^{\rm bkg} &=& \frac{0.16 \epsilon^{-1.1}}{1+11 \epsilon^{0.9}+3.2\epsilon^{2.15}}
{\rm GeV^{-1} cm^{-2} s^{-1} sr^{-1}}\nonumber\\
\Phi_{\rm sec,\; e^-}^{\rm bkg} &=& \frac{0.70\epsilon^{0.7}}{1+110\epsilon^{1.5}+600 \epsilon^{2.9}
	+580\epsilon^{4.2}} {\rm GeV^{-1} cm^{-2} s^{-1} sr^{-1}}\nonumber\\
\Phi_{\rm sec,\; e^+}^{\rm bkg} &=& \frac{4.5 \epsilon^{0.7}}{1+650 \epsilon^{2.3}+1500 \epsilon^{4.2}}
{\rm GeV^{-1} cm^{-2} s^{-1} sr^{-1} }\,\label{eq:bkg}
\end{eqnarray}
where the dimensionless parameter $\epsilon=E$/(1 GeV). We have used these background fluxes for obtaining constraints with HAWC, GRAPES-3, CASA-MIA and H.E.S.S. data sets. For the CALET, AMS-02, and DAMPE datasets, we used the $e^+e^-$ background flux which is given in~\cite{CALET:2023emo}.

\section{Result and discussions}\label{sec:result}

We have performed chi-square analysis to obtain the upper bound on the annihilation cross-section and lower limit on the DM lifetime. We define the chi-square as
\begin{eqnarray}
    \chi^2=\sum \frac{\left(\rm data-(model+background)\right)^2}{\sigma^2},
\end{eqnarray}
where `data' represents the observed flux at different experiments, `model' indicates the flux from the DM annihilation or decay, `background' is given in Eq.(\ref{eq:bkg}), and $\sigma$ denotes the error bar in the observed flux. Minimum of the chi-square, $\chi^2_{\rm min}$ is calculated by varying $\langle\sigma v\rangle$ or $\tau_{\rm DM}$ for a particular mass of DM, $M_{\rm DM}$. After getting the $\chi^2_{\rm min}$, the 95\% C.L. upper limit is obtained by using the equation $\chi^2-\chi^2_{\rm min}\approx2.71$. 

As for the HAWC, CASA-MIA, and GRAPES-3, only upper limits on the flux are available rather than the flux measurements, we do not perform $\chi^2$ fit for these experiments. We impose the condition that the predicted flux does not exceed the experimental upper limits in any energy bin, $i$, as
\begin{eqnarray}
    \Phi^{\rm DM}_i<\Phi_i^{\rm UL},
\end{eqnarray}
where $\Phi_i^{\rm UL}$ is the upper limits on the flux and $\Phi^{\rm DM}_i$ is the DM induced flux plus astrophysical background, Eq. (\ref{eq:bkg}).

\begin{figure}[h]
    \centering
    \includegraphics[scale=0.3]{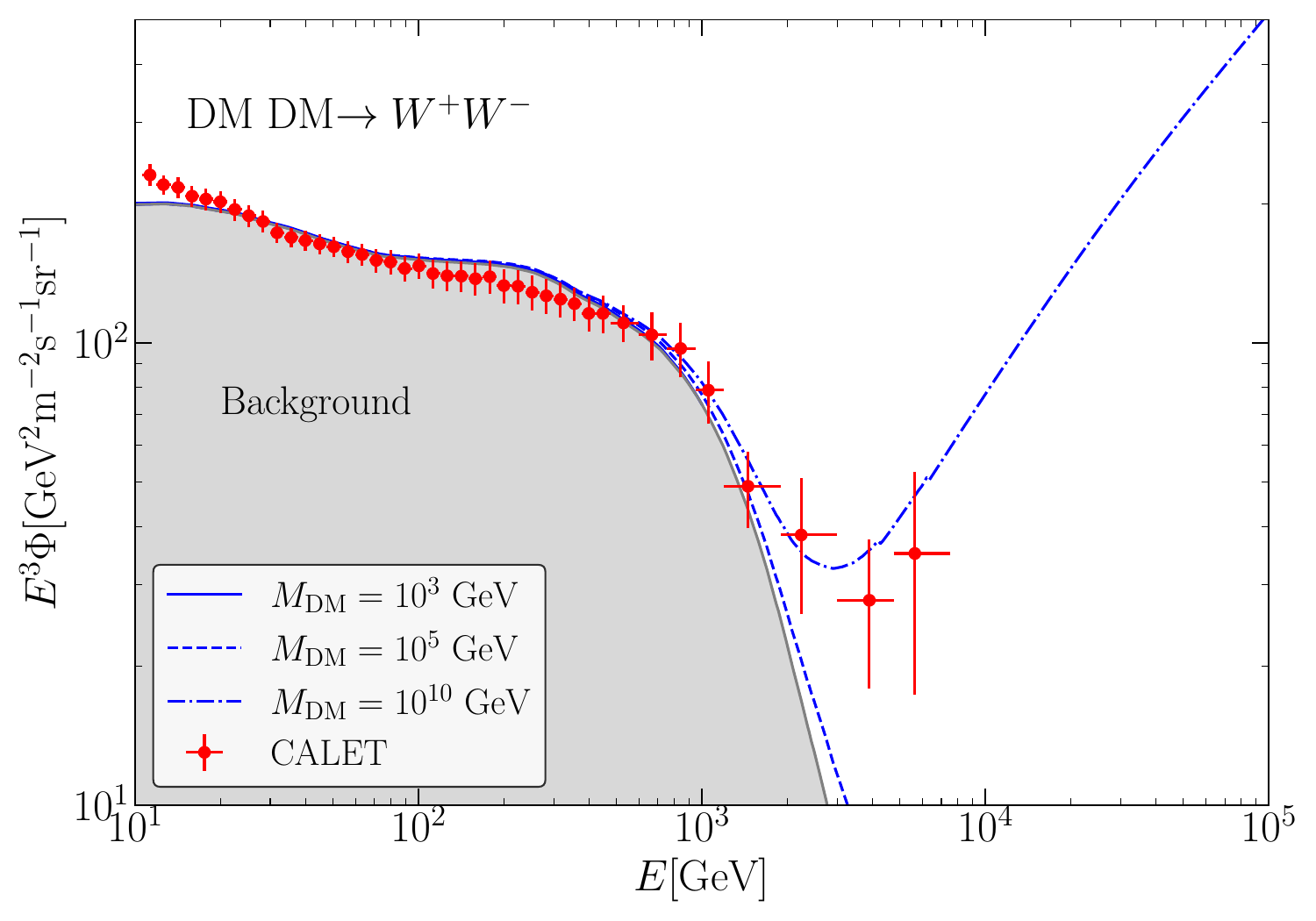}
    \includegraphics[scale=0.3]{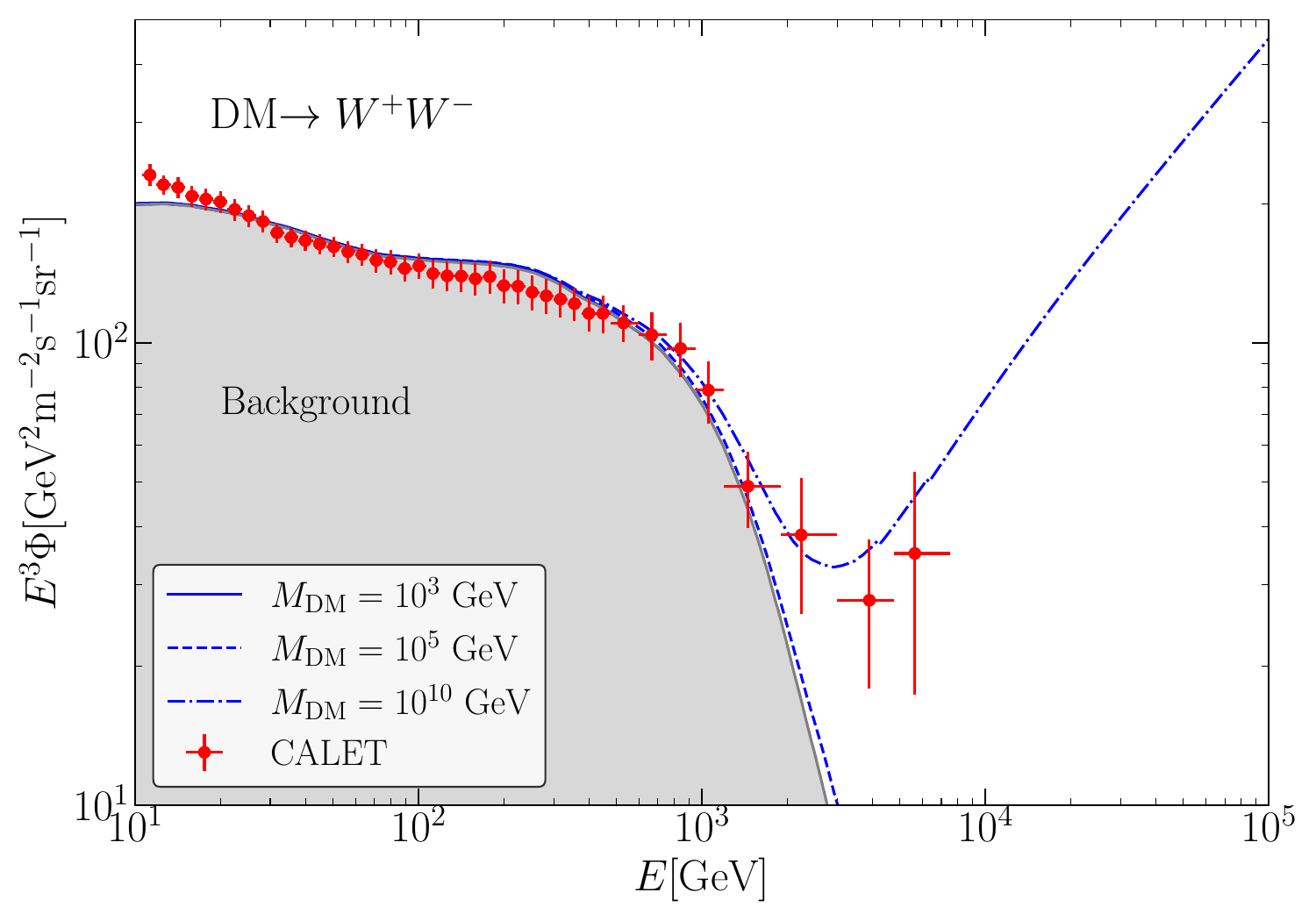}\caption{\textit{Left:} Total electron flux ($e^-+e^+$) from DM annihilation into the $W^+W^-$ channel for $M_{\rm DM}=10^3$ GeV (blue solid), $10^5$ GeV (blue dashed), and $10^{10}$ GeV (blue dash-dotted). The curves correspond to the annihilation cross-sections at the 95\% C.L. limit, with values of $1.268\times10^{-24}~\rm cm^3/s$, $3.8339\times10^{-22}~\rm cm^3/s$, and $1.7046\times10^{-13}~\rm cm^3/s$ for these respective masses. The red cross represents the CALET electron flux data~\cite{CALET:2023emo}. The gray-shaded region is the all-electron background, including secondaries, distant supernovae, and electron flux from all pulsars, taken from~\cite{CALET:2023emo}. \textit{Right:} Total electron flux from DM decay into the $W^+W^-$ channel for $M_{\rm DM}=10^3$ GeV (blue solid), $10^5$ GeV (blue dashed), and $10^{10}$ GeV (blue dash-dotted). The curves correspond to the decay lifetime at the 95\% C.L. limit, with values of $2.0217\times10^{27}~\rm s$, $1.0751\times10^{27}~\rm s$, and $2.747\times10^{23}~\rm s$ for these respective masses.}
    \label{fig:flux}
\end{figure}

In the \textit{left} panel of Fig.~\ref{fig:flux}, we present the all electron flux (\textit{i.e.,} $e^-+e^+$) originating from the DM annihilating to $W^+W^-$ channel for three representative benchmark masses, $M_{\rm DM}=\{10^3,\ 10^5,\ 10^{10}\}$ GeV. Considering the CALET all electron flux, we compute the 95\% C.L. upper limit on $\langle\sigma v\rangle$ for these masses and are given as $\langle\sigma v\rangle=1.268\times10^{-24}\rm cm^3/s$ for $M_{\rm DM}=10^{3}$ GeV, $\langle\sigma v\rangle=3.8339\times10^{-21}\rm cm^3/s$ for $M_{\rm DM}=10^{5}$ GeV, and $\langle\sigma v\rangle=1.7046\times10^{-13}\rm cm^3/s$ for $M_{\rm DM}=10^{10}$ GeV. The CALET flux is shown with the red crosses. The gray curve represents the astrophysical background~\cite{CALET:2023emo}. The blue solid line corresponds to the all-electron flux from DM annihilation plus astrophysical background for $M_{\rm DM}=10^3$ GeV. The same is shown for $M_{\rm DM}=10^5$ GeV and $M_{\rm DM}=10^{10}$ GeV with blue dashed and blue dashed-dotted lines, respectively. In the \textit{right} panel of Fig.~\ref{fig:flux}, we show the all-electron flux produced from the decay of DM with the same masses and the channel as mentioned for the annihilation case. The 95\% C.L. limit on the lifetime for these masses are $\tau_{\rm DM}=2.0217\times10^{27}$s, $\tau_{\rm DM}=1.0751\times10^{27}$s, and $\tau_{\rm DM}=2.747\times10^{23}$s.

 \begin{figure}[h]
	\centering	\includegraphics[scale=0.3]{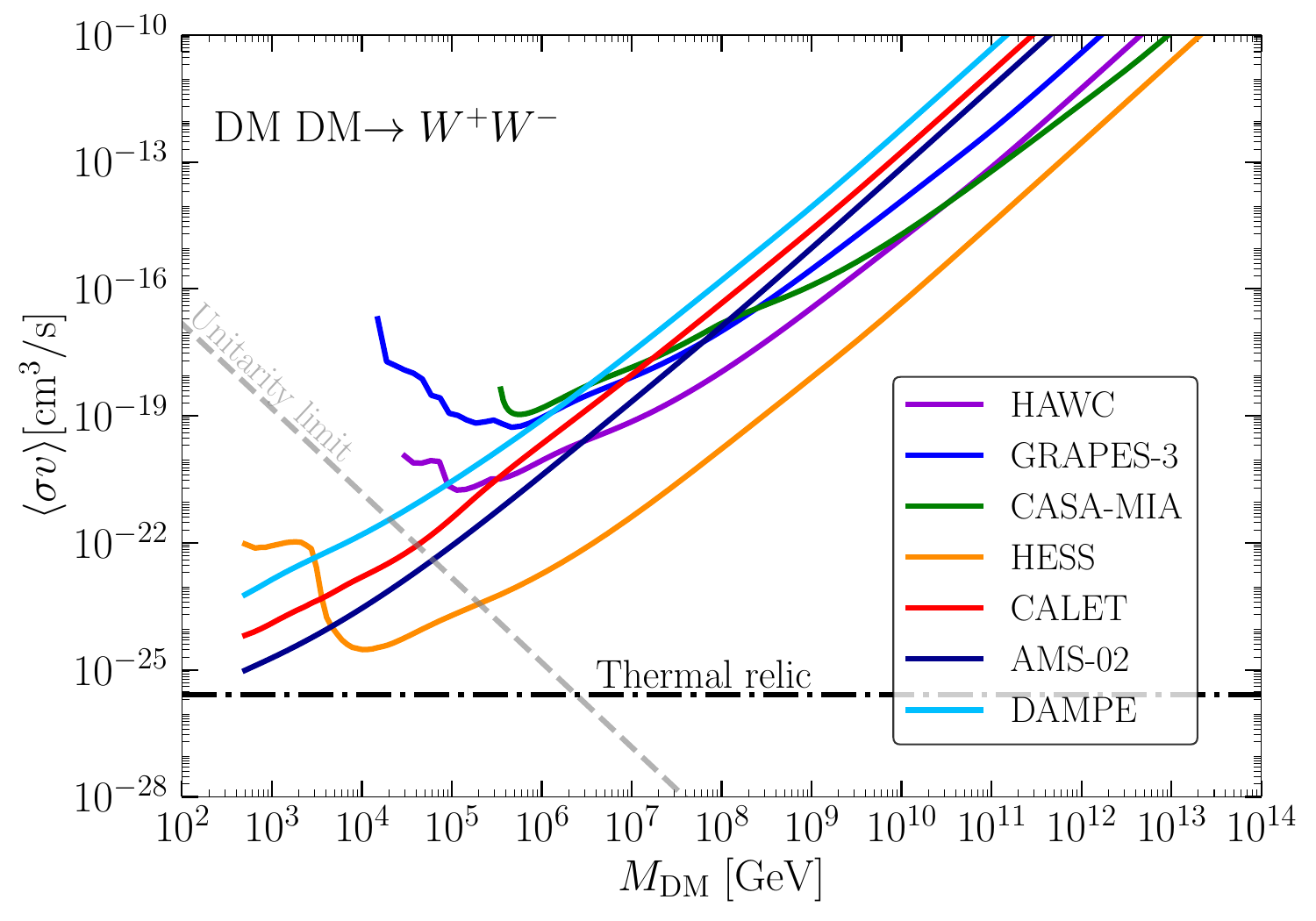}
    \includegraphics[scale=0.3]{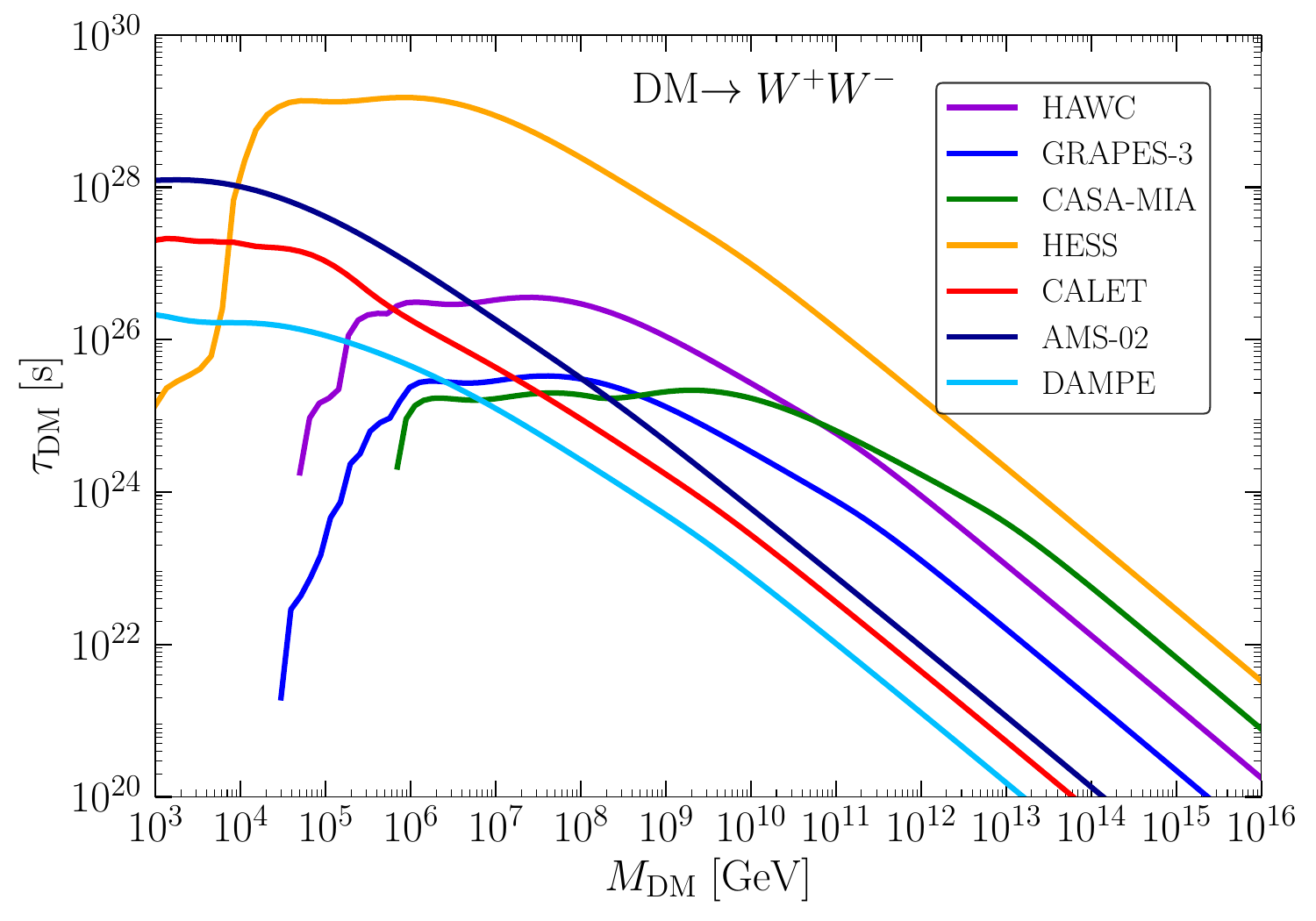}
	\caption{95\% C.L. limits on DM annihilation cross-section (\textit{left}) and DM lifetime (\textit{right}) from CALET, AMS-02, HAWC, GRAPES-3, CASA-MIA, DAMPE, and H.E.S.S. for $W^+W^-$ channel. The black dashed-dotted horizontal line represent the approximate value of the canonical of the annihilation cross-section corresponding to the \textquotedblleft WIMP miracle\textquotedblright, $\langle\sigma v\rangle=3\times10^{-26}~\rm cm^3/s$~\cite{Steigman:2012nb,Saikawa:2020swg}. The gray dashed line depict the unitarity limit, see text for more details.}
	\label{fig:dmlimitsww}
\end{figure}

In Fig.~\ref{fig:dmlimitsww}, we show the 95\% C.L. limits on DM annihilation cross-section and lifetime for the $W^+W^-$ channel. The \textit{left} panel shows the upper bounds on the thermally averaged annihilation cross-section $\langle\sigma v\rangle$ as a function of the DM mass $M_{\rm DM}$ using 
HAWC, GRAPES-3, CASA-MIA, H.E.S.S., CALET, and AMS-02 data. In this plot, we show the reference value of the canonical cross-section for weakly interacting massive particle (WIMP) models and also the unitarity bound. Note that the unitarity bounds are not always applicable to models of our considerations; however, we show the line as a reference. Across the mass range $500~{\rm GeV}\leq M_{\rm DM}\leq10^{14}~{\rm GeV}$, the experiments set stringent bounds, excluding cross-sections above $\sim10^{-25}~\rm cm^3/s$ at low masses and tightening to $\mathcal{O}(10^{-24}){~\rm cm^3/s}-\mathcal{O}(10^{-20}){~\rm cm^3/s}$ for multi-TeV and multi-PeV-scale DM. Up to $M_{\rm DM}\sim10^4$ GeV, AMS-02 provides the most stringent constraints. The HESS limit dominates for $M_{\rm DM}\sim10^4~{\rm GeV}-10^{14}~{\rm GeV}$. The \textit{right} panel presents the corresponding lower limits on the DM lifetime $\tau_{\rm DM}$ for the same decay channel. Here, the constraints span the mass range $10^3~{\rm GeV}-10^{16}~{\rm GeV}$. In the low mass range up to $\sim\mathcal{O}(10^4)~{\rm GeV}$, the AMS-02 imposes the strongest constraint. For masses around $10^4~{\rm GeV}-10^{16}~{\rm GeV}$, HESS provides the most stringent limits, reaching $\tau_{\rm DM}\gtrsim10^{29}~\rm s$ for $M_{\rm DM}\sim10^6$ GeV.

\begin{figure}[h]
	\centering	\includegraphics[scale=0.3]{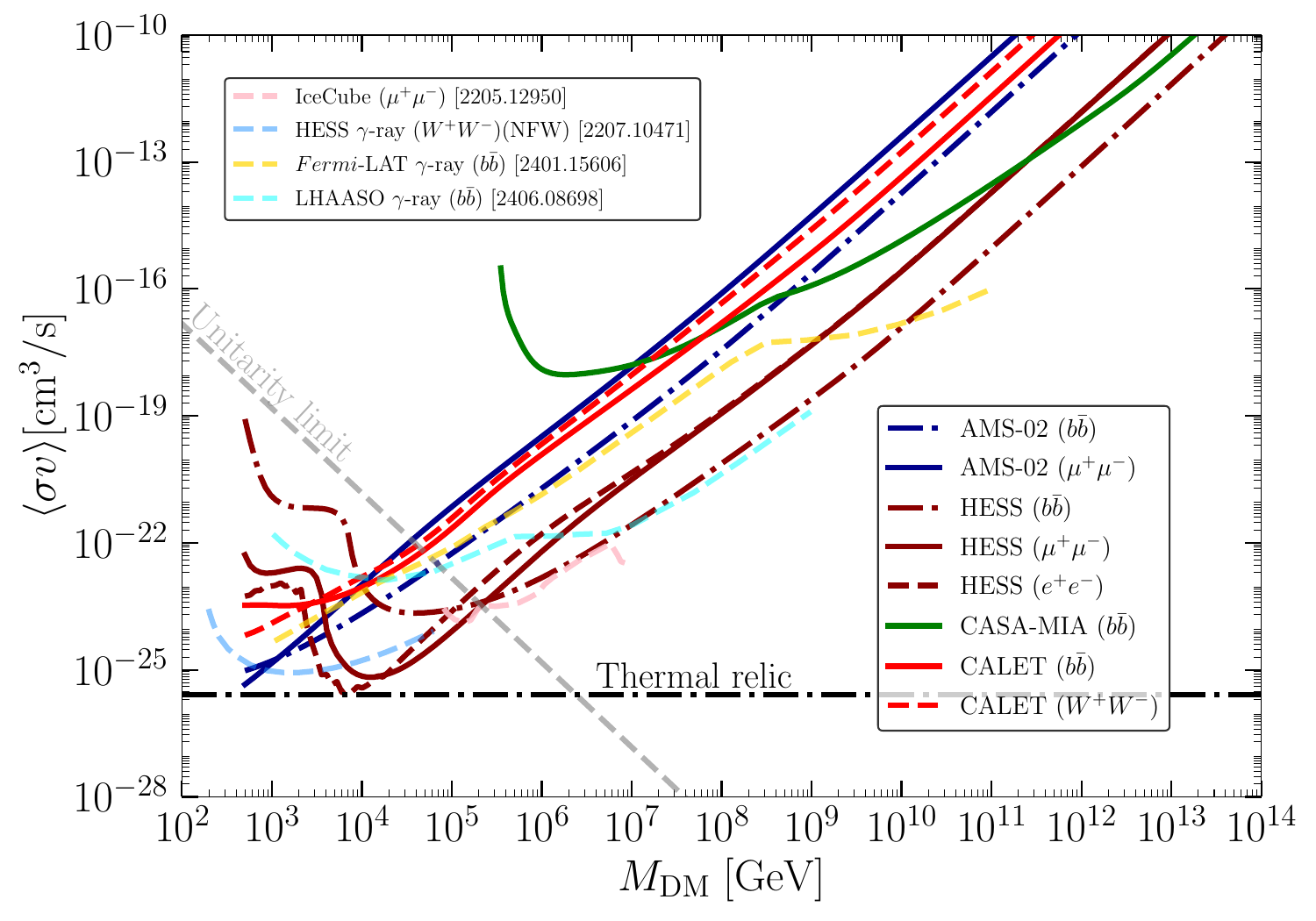}
    \includegraphics[scale=0.3]{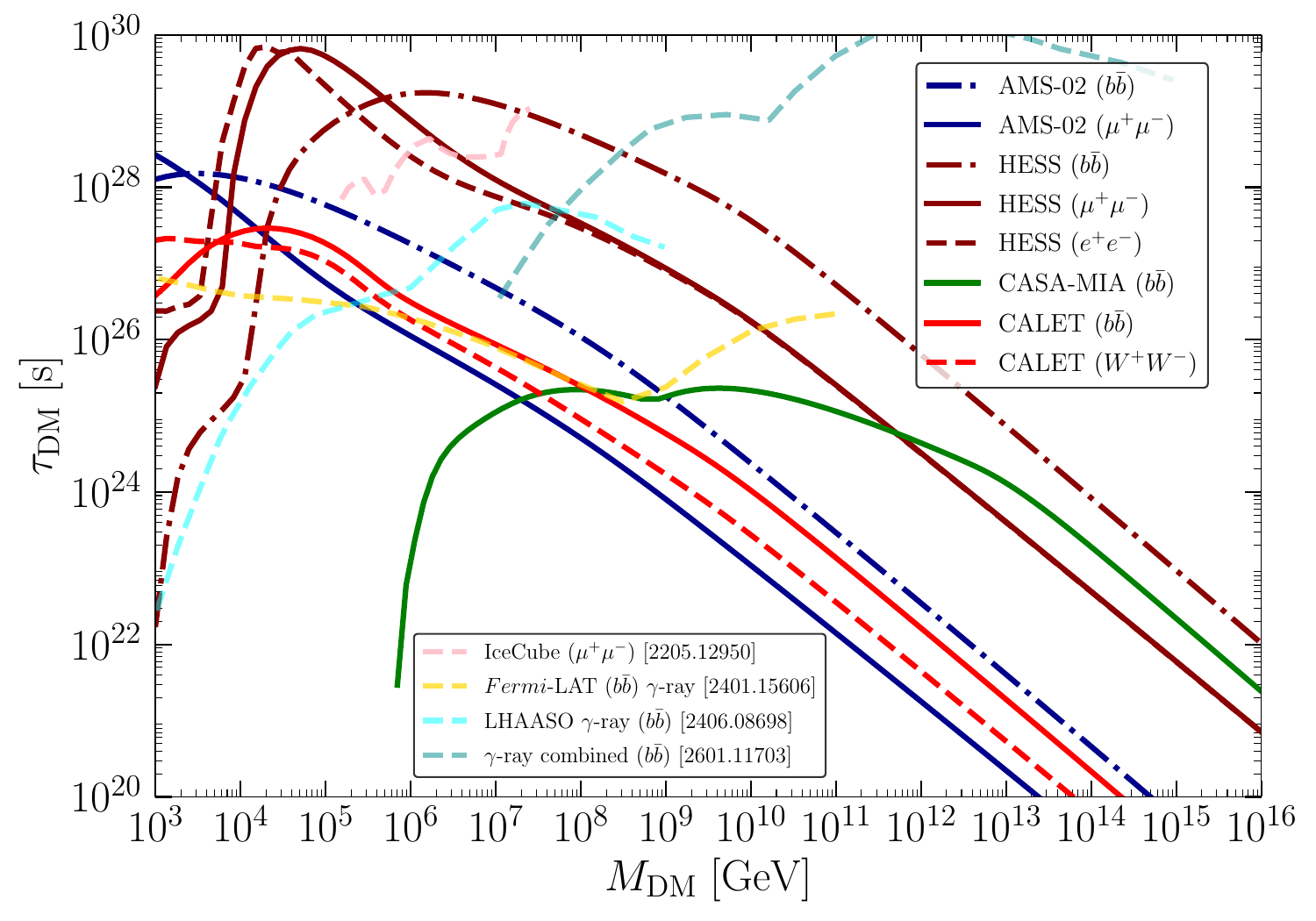}
	\caption{\textit{Left:} Summary of 95\% C.L. limits on DM annihilation cross-section, $\langle\sigma v\rangle$ as a function of DM mass, $M_{\rm DM}$ for AMS-02, CALET, HESS, and CASA-MIA are shown. \textit{Right:} Summary of 95\% C.L. limits on DM lifetime, $\tau_{\rm DM}$ as a function of DM mass, $M_{\rm DM}$ for AMS-02, CALET, HESS, and CASA-MIA are shown. Constraints from neutrino telescope: IceCube \cite{IceCube:2022clp} and gamma-ray observatories: HESS \cite{HESS:2022ygk}, Fermi-LAT\cite{Song:2024vdc}, LHAASO\cite{LHAASO:2024upb},and the combined constraints \cite{Chianese:2026cfz} from KASCADE \cite{KASCADE:2005ynk}, KASCADE-Grande \cite{KASCADEGrande:2017vwf}, Telescrope Array \cite{TelescopeArray:2018rbt}, Pierre Auger \cite{PierreAuger:2025jwt},  are also shown for comparison.}
	\label{fig:summaryplot}
\end{figure}

We summarize our results in Fig.~\ref{fig:summaryplot} for the most stringent annihilation and decay modes. The constraints from the $e^+e^-$, $\mu^+\mu^-$, $\tau^+\tau^-$, $b\bar{b}$ modes are provided in Appendix \ref{app:otherchannel}. In the \textit{left} panel of Fig.~\ref{fig:summaryplot}, we show the 95\% C.L. limits of AMS-02 ($\mu^+\mu^-$, $b\bar{b}$), HESS ($e^+e^-$, $\mu^+\mu^-$, $b\bar{b}$), CASA-MIA ($b\bar{b}$), and CALET ($W^+W^-$, $b\bar{b}$). The AMS-02 constraint is stronger in the mass range $M_{\rm DM}\sim500~{\rm GeV}-2~{\rm TeV}$. The HESS sets the most stringent limit up to $M_{\rm DM}\sim10^{14}$ GeV. The combined lower limits on the DM lifetime, $\tau_{\rm DM}$, are presented in the \textit{right} panel of Fig.\ref{fig:summaryplot}. Here, AMS-02 leads the bound in the mass range $M_{\rm DM}\sim10^3~{\rm GeV}-10^4~{\rm GeV}$, HESS is relevant in the mass range $10^4~{\rm GeV}-10^{16}~{\rm GeV}$. In Fig. \ref{fig:summaryplot}, we also show the limits from neutrino telescope: IceCube \cite{IceCube:2022clp} and gamma-ray observatories: HESS \cite{HESS:2022ygk}, Fermi-LAT\cite{Song:2024vdc}, LHAASO\cite{LHAASO:2024upb}, and the combined constraints \cite{Chianese:2026cfz} from KASCADE \cite{KASCADE:2005ynk}, KASCADE-Grande \cite{KASCADEGrande:2017vwf}, Telescrope Array \cite{TelescopeArray:2018rbt}, Pierre Auger \cite{PierreAuger:2025jwt}, with various dashed colored lines as indicated in the figure inset for comparison purpose. For a decaying DM, we see that in the low to intermediate mass region ($M_{\rm DM}\lesssim10^9$ GeV), the $e^+e^-$ constraint is much stronger, while for the larger DM masses, $\gamma$-ray data lead the constraints. On the other hand, for the annihilating DM case, $\gamma$-ray data give a stringent constraint up to 3 TeV mass of DM. $e^+e^-$ constraints are the strongest in the mass range 3 TeV to 100 TeV. Beyond this range, the neutrino and $\gamma$-ray data remain comparable to the $e^+e^-$ limits up to $M_{\rm DM}\sim10^{11}$ GeV. The $e^+e^-$ constraints lead the limit for $M_{\rm DM}\gtrsim10^{11}$ GeV.

\section{Conclusions and future outlook}\label{sec:conlc}
In this work, we have studied the constraints on heavy dark matter in the mass range $500~{\rm GeV}\leq M_{\rm DM}\leq10^{16}~{\rm GeV}$ combining all-electron flux from various experiments. Assuming two-body SM final states, $W^+W^-$, $b\bar{b}$, $\mu^+\mu^-$, $\tau^+\tau^-$, and $e^+e^-$, we derived limits on both the annihilation cross-section and the decay lifetime using data from CALET, AMS-02, H.E.S.S., HAWC, GRAPES-3, DAMPE, and CASA-MIA.  We find that in the low-mass regime up to 2 TeV, AMS-02 places stringent constraints on $\langle\sigma v\rangle$. For unstable dark matter, AMS-02 provides strong limits on the lifetime, excluding $\tau_{\rm DM}({\rm DM}\rightarrow b\bar{b}) \lesssim \mathcal{O}(10^{28})$~s for $M_{\rm DM}=1$ TeV. In the intermediate to high-mass regime, typically $10^{4}~{\rm GeV} \lesssim M_{\rm DM} \lesssim 10^{16}~{\rm GeV}$, HESS yields the most stringent constraints on both $\langle\sigma v\rangle$ and $\tau_{\rm DM}$ from $e^+e^-$ flux measurements. The constraints from $e^+e^-$ data remain competitive with those from $\gamma$-ray and neutrino data. The next generation of cosmic-ray and gamma-ray observatories will significantly extend the sensitivity to dark matter signatures across a wide mass range. Improved energy resolution, larger effective areas, and longer observation times will allow these facilities to either detect signals of DM decay or annihilation or push the exclusion limits on $\langle\sigma v\rangle$ and $\tau_{\rm DM}$ even further.

\noindent
\acknowledgments
K.K. would like to thank Susumu Inoue for fruitful discussion. P.K.P. would like to acknowledge the Ministry of Education, Government of India, for providing financial support for his research via the Prime Minister’s Research Fellowship (PMRF) scheme. This work was also supported in part by the MEXT Leading Initiative for Excellent Young Researchers Grant Number 2023L0013 (N.H), the JSPS KAKENHI Grant Nos. JP22K14035 (N.H), JP23KF0289 (K.K.), and JP24K07027 (K.K.), and the MEXT KAKENHI Grants No. JP24H01825 (K.K.).

\appendix
\section{Gamma-ray and $e^\pm$ fluxes from different experiments}\label{app:data}
In the analysis, we have used $e^-+e^-$ data from CALET, DAMPE, and HESS; $e^+$ data from AMS-02; and $\gamma$-ray flux from HAWC, GRAPES-3, and CASAMIA.
\begin{table}[h]
\begin{center}
		\begin{tblr}{
					colspec={|l|l|},
					 row{1}={font=\bfseries}
				}
				\toprule Data set & Source\\
				\toprule
				CALET $e^++e^-$ flux & Fig. 4 in \cite{CALET:2023emo}.\\
                \hline
				DAMPE $e^++e^-$ flux& Fig. 2 in \cite{Alemanno:2022eeb}.\\
                \hline
				HESS $e^++e^-$ flux& Fig. 3 in \cite{HESS:2024etj}. \\\hline
				AMS-02 $e^+$ flux& Fig. 34 in \cite{AMS:2021nhj}. \\
                \hline
				HAWC $\gamma$-ray flux& Fig. 3 in \cite{HAWC:2022uka}. \\
                \hline
				GRAPES-3 $\gamma$-ray flux& Fig. 3 in \cite{HAWC:2022uka}. \\
                \hline
				CASA-MIA $\gamma$-ray flux& Fig. 3 in \cite{HAWC:2022uka}. \\
				\bottomrule
		\end{tblr}
		\caption{Sources of the experimental flux used in deriving limits on $\langle\sigma{v}\rangle$ and $\tau_{\rm DM}$ as a function of DM mass.}
		\label{tab:tab1}
\end{center}
\end{table}
We have listed the sources of the data in the Table. \ref{tab:tab1}.

\section{Effect of propagation model uncertainty on $\langle\sigma{v}\rangle/\tau_{\rm DM}$ limits}\label{app:flux}

\begin{figure}[h]
    \centering    \includegraphics[scale=0.3]{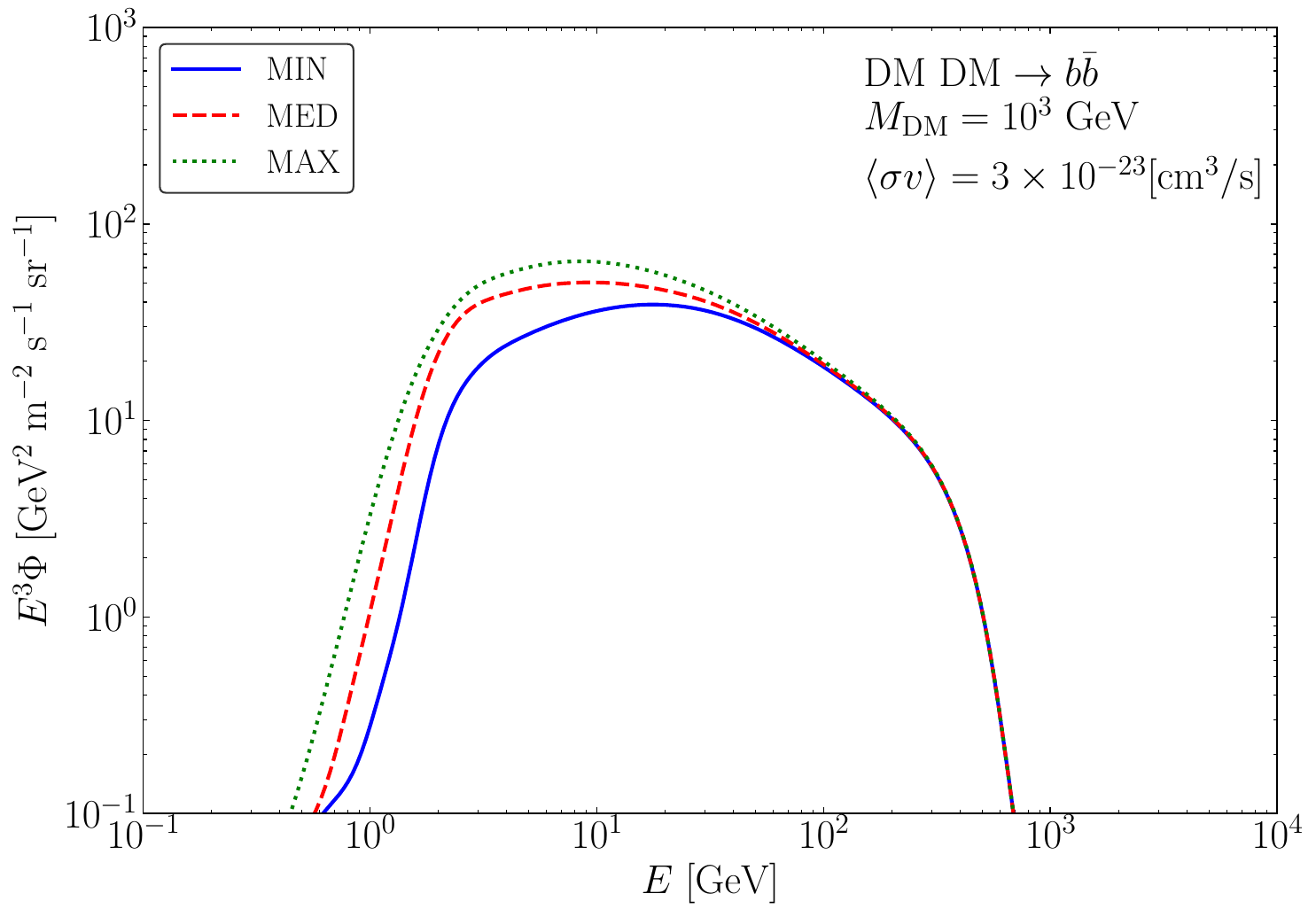}
    \includegraphics[scale=0.3]{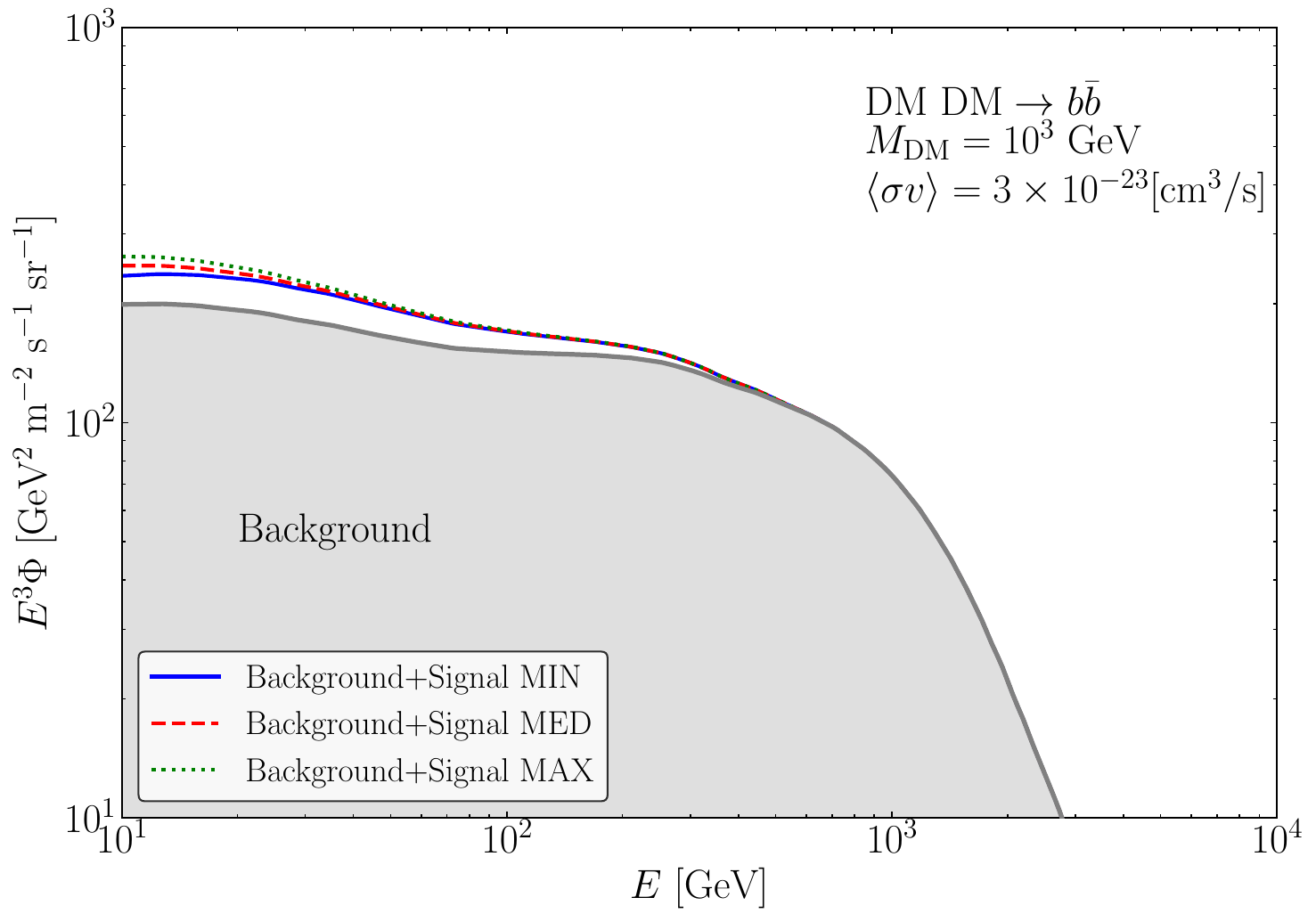}
    \caption{\textit{Left:} $e^\pm$ flux as a function of energy originating from ${\rm DM~DM}\rightarrow b\bar{b}$ process for $M_{\rm DM}=10^3$ GeV and $\langle\sigma{v}\rangle=3\times10^{-23}~ \rm cm^3/s$ for \texttt{MIN}, \texttt{MED}, and \texttt{MAX} propagation models. \textit{Right:} $e^\pm$ flux + astrophysical background as a function of energy for the same parameters as of the \textit{left} panel.}
    \label{fig:flux_comp}
\end{figure}

The left panel of Fig. \ref{fig:flux_comp} shows the DM-induced $e^\pm$ flux alone (without the  astrophysical background) for DM DM$\to b\bar{b}$ at $M_\mathrm{DM}=10^3$~GeV and $\langle\sigma v\rangle = 3\times10^{-23}$ cm$^3$/s. At low energies ($E \ll M_\mathrm{DM}$), the three propagation models yield fluxes that differ by a factor of $\sim2$--$3$. This spread arises because low-energy positrons have long propagation lengths and therefore sample the full extent of the diffusion halo; their flux is consequently sensitive to the diffusive halo height $L$ and the diffusion coefficient $K_0$, which are the primary quantities that differ between the {\tt MIN}, {\tt MED}, and {\tt MAX} parameter sets. However, the three curves converge as $E\to M_\mathrm{DM}$. This convergence is a direct consequence of the rapid energy losses that high-energy positrons suffer.

The right panel of Fig. \ref{fig:flux_comp} shows the total flux (astrophysical background plus DM signal) for the same three propagation models. The three curves are nearly indistinguishable across the entire energy range, because (i) the DM signal is small compared to the background at low energies, where the propagation spread is largest, and (ii) at high energies near $E \sim M_\mathrm{DM}$, the signal flux drops sharply toward the kinematic endpoint and becomes negligible relative to the background, so the propagation model choice is irrelevant there.

Although the DM signal flux differs by a factor of $\sim 2$--$3$ between the {\tt MIN}, {\tt MED}, and {\tt MAX} propagation models, it remains subdominant relative to the astrophysical background across the entire relevant energy range. Consequently, the fractional uncertainty on the total predicted flux, signal plus background, is reduced to only a few percent. This directly translates into an uncertainty of less than $\sim 10\%$ on the inferred upper limit on $\langle\sigma v\rangle$ (or lower limit on $\tau_\mathrm{DM}$) across propagation models. This conclusion holds for all DM masses and annihilation channels considered in this work.

\section{Fitted flux for different experiments}\label{app:flux0}
In this section, we present the fitted fluxes for different experiments across two choices of DM masses for the $\mu^+\mu^-$ channel.

\subsection{CALET}

\begin{figure}[H]
\centering
\includegraphics[scale=0.3]{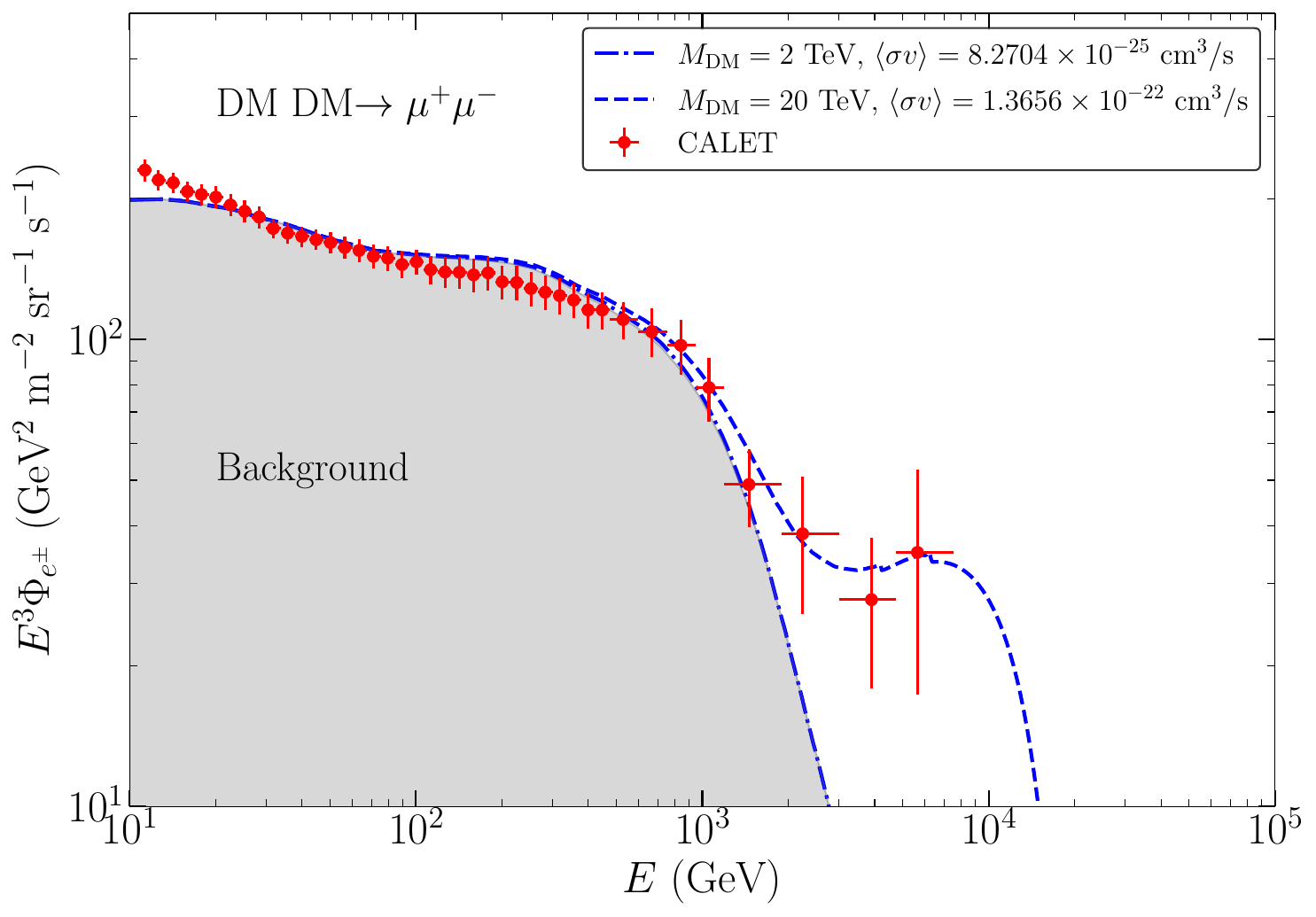}
\includegraphics[scale=0.3]{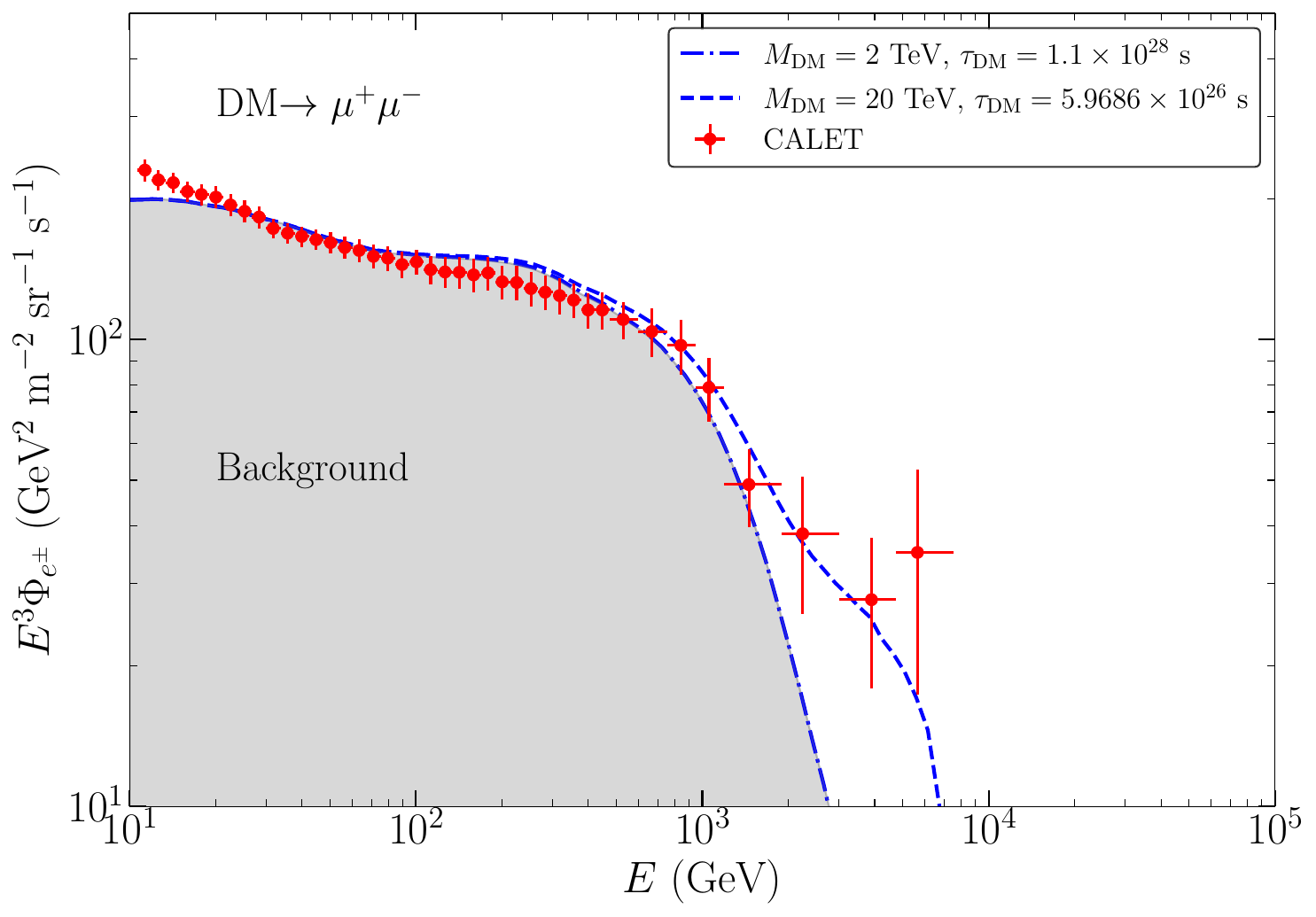}
\caption{\textit{Left:} Total electron flux ($e^-+e^+$) from DM annihilation into the $\mu^+\mu^-$ channel for $M_{\rm DM}=2$ TeV (blue dash-dotted) and $20$ TeV (blue dashed). The curves correspond to the annihilation cross-sections at the 95\% C.L. limit as mentioned in the figure inset. The red cross represents the CALET electron flux data~\cite{CALET:2023emo}. The gray-shaded region is the all-electron background, including secondaries, distant supernovae, and electron flux from all pulsars, taken from~\cite{CALET:2023emo}. \textit{Right:} Total electron flux from DM decay into the $\mu^+\mu^-$ channel for the same DM masses. The curves correspond to the decay lifetime at the 95\% C.L. limit.}
\label{fig:caletflux}
\end{figure}

In Fig. \ref{fig:caletflux}, we show the flux as a function of energy for two choices of DM masses for the CALET experiment.

\subsection{AMS-02}

\begin{figure}[H]
\centering
\includegraphics[scale=0.3]{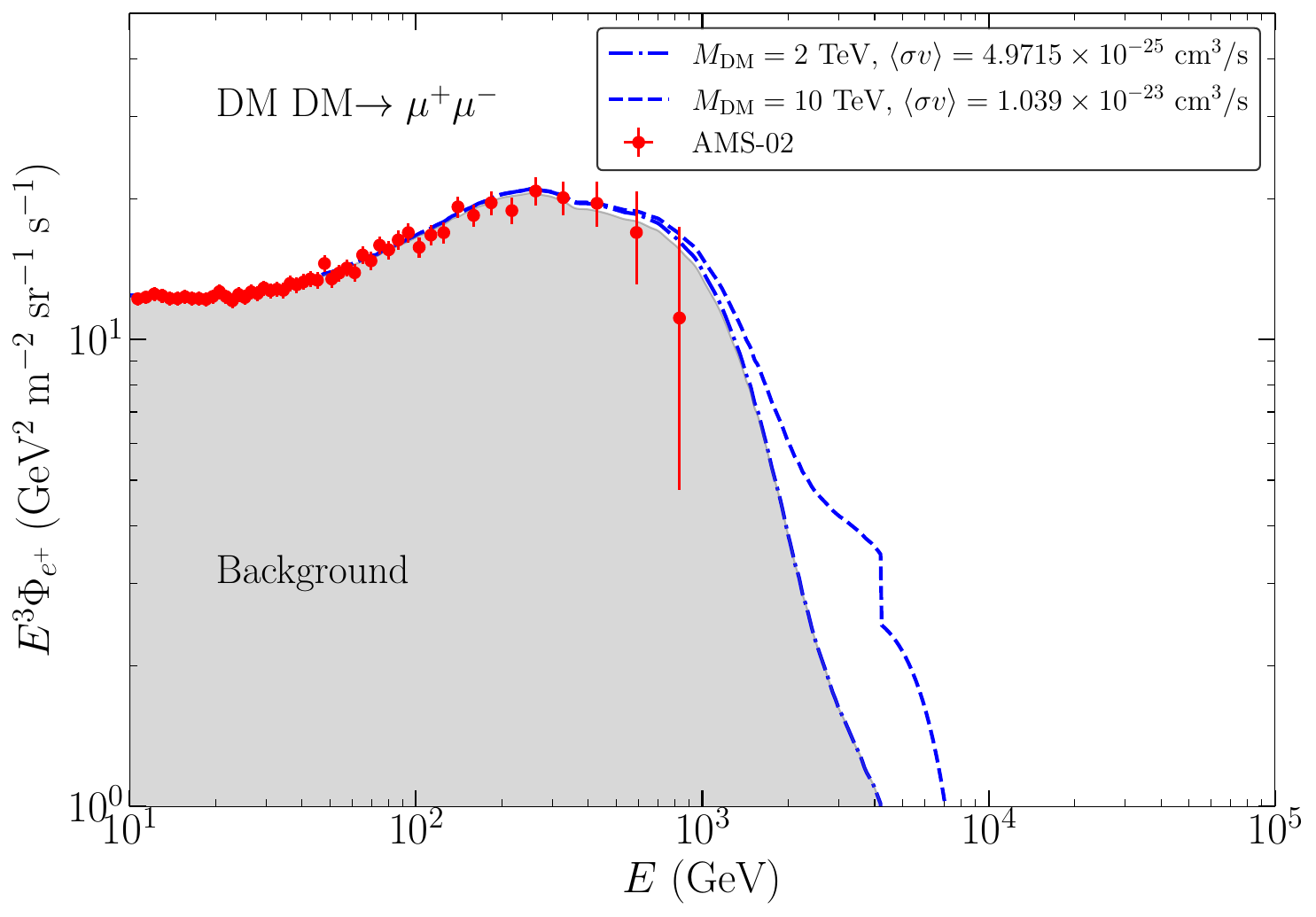}
\includegraphics[scale=0.3]{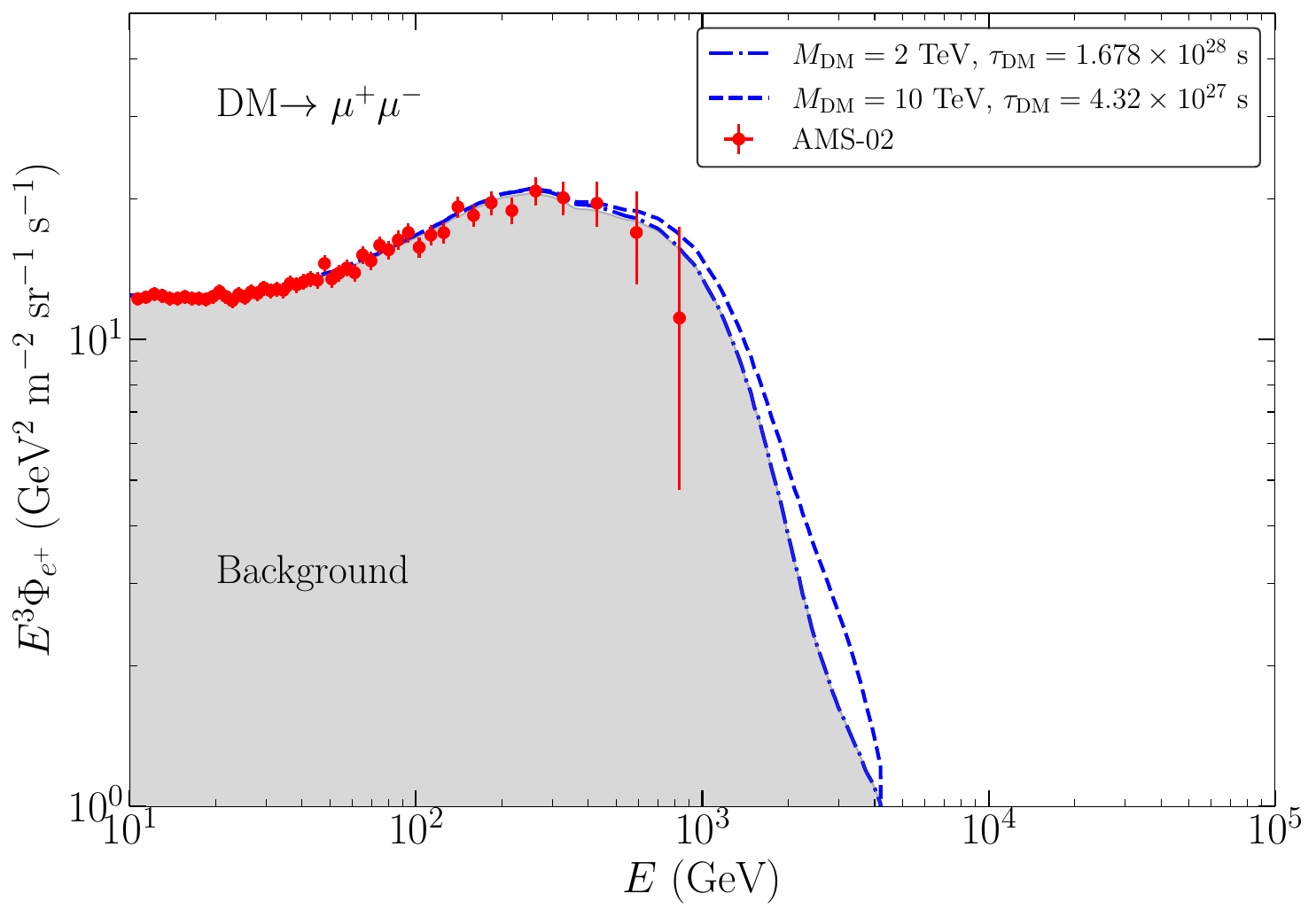}\caption{Same as in Fig. \ref{fig:caletflux} but for the AMS-02 positron flux.}
\label{fig:ams2flux}
\end{figure}

In Fig. \ref{fig:ams2flux}, we show the flux as a function of energy for two choices of DM masses for the AMS-02 positron flux measurements.

\subsection{DAMPE}

\begin{figure}[H]
\centering
\includegraphics[scale=0.3]{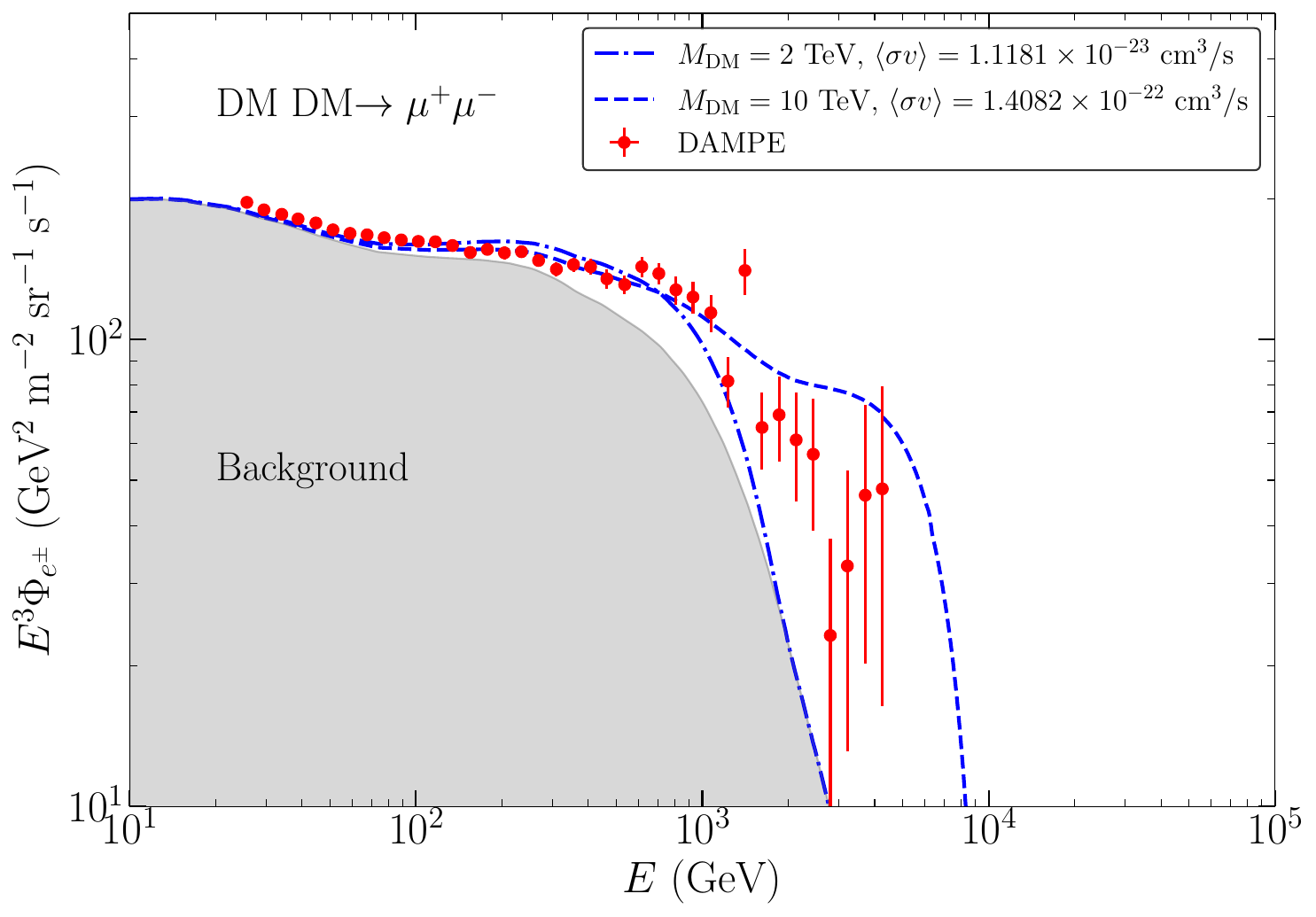}
\includegraphics[scale=0.3]{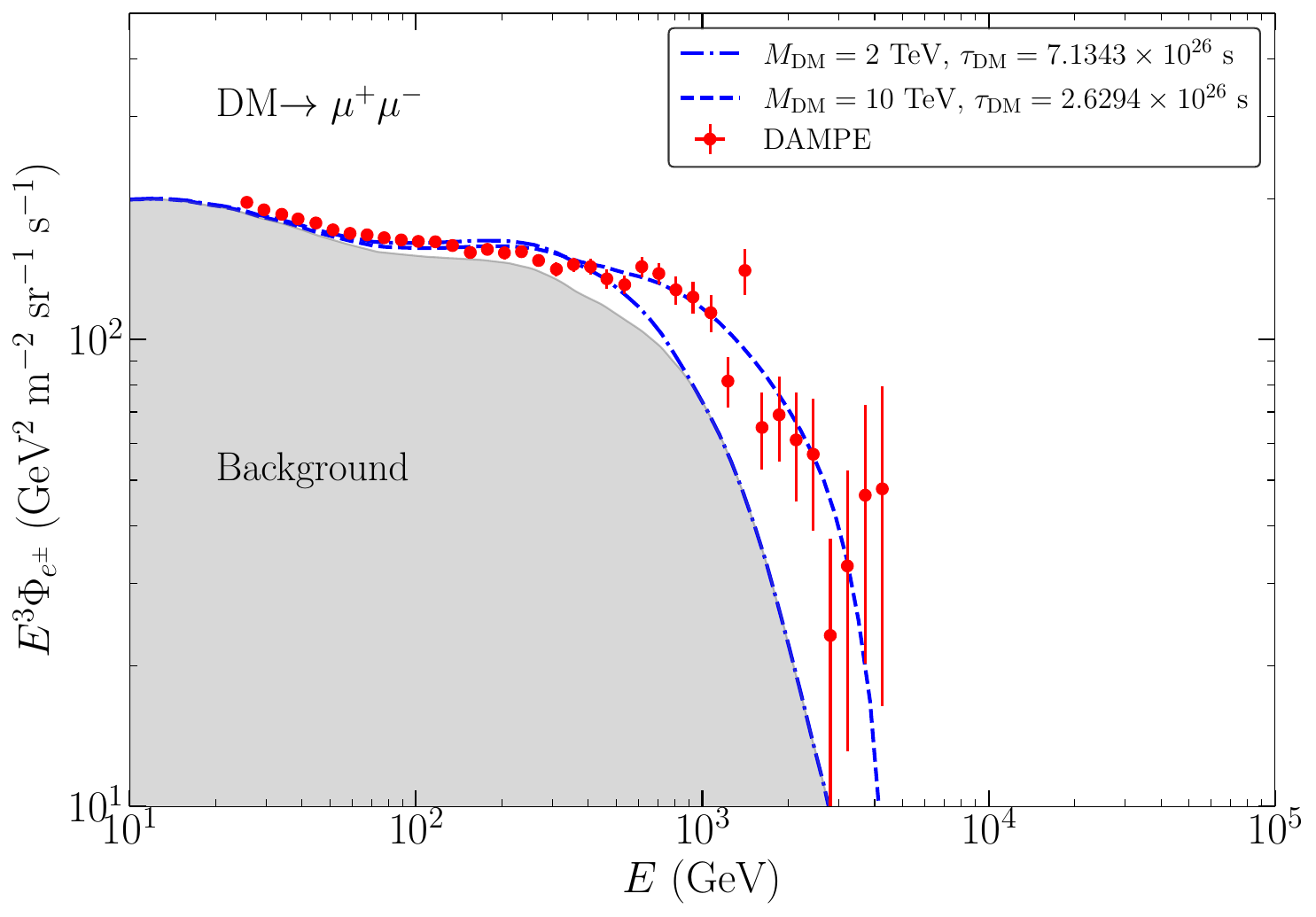}\caption{Same as in Fig. \ref{fig:caletflux} but for the DAMPE electron-positron flux.}
\label{fig:dampeflux}
\end{figure}

In Fig. \ref{fig:dampeflux}, we show the flux as a function of energy for two choices of DM masses for the DAMPE electron-positron flux measurements.

\subsection{HESS}

\begin{figure}[h]
\centering
\includegraphics[scale=0.3]{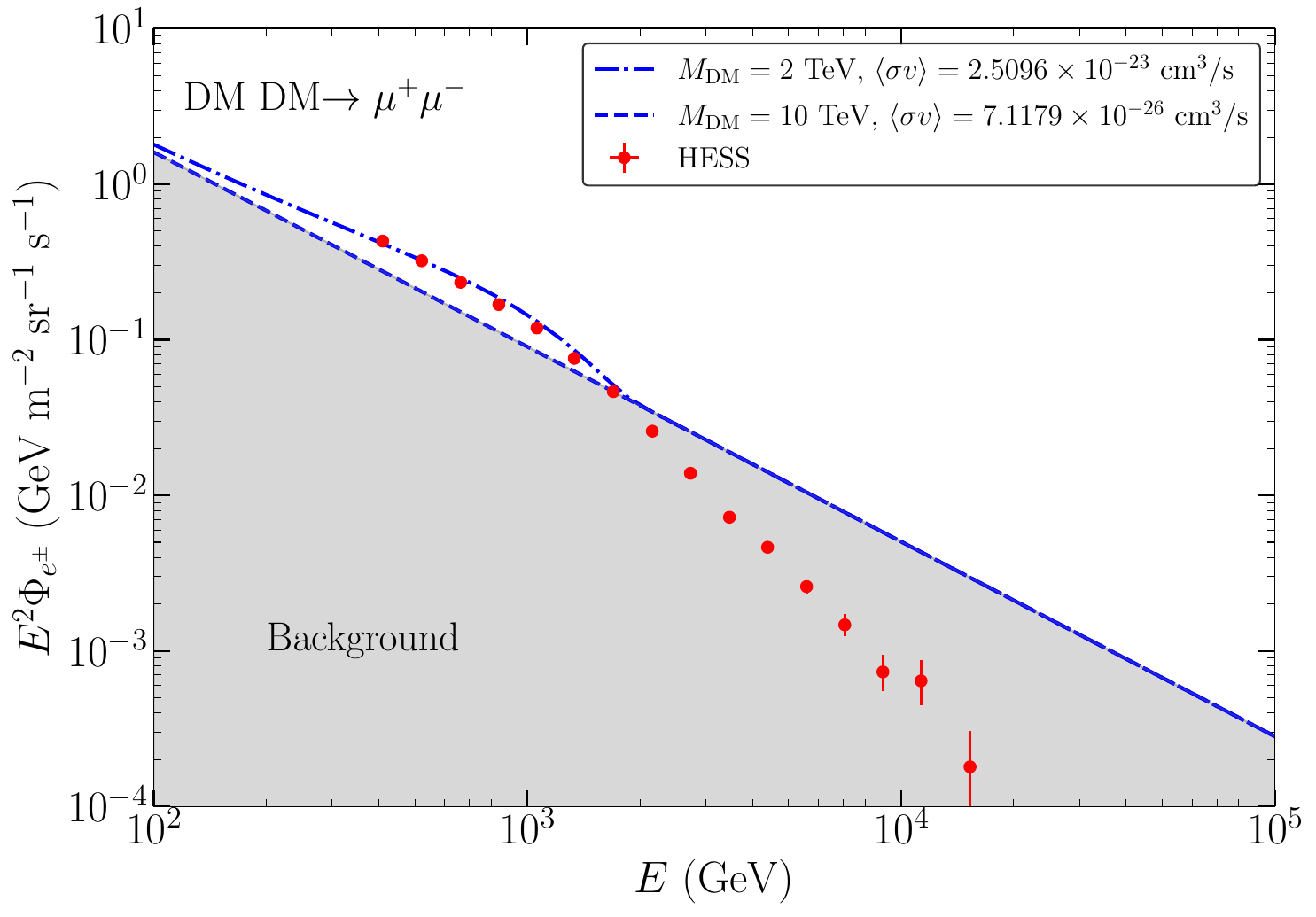}
\includegraphics[scale=0.3]{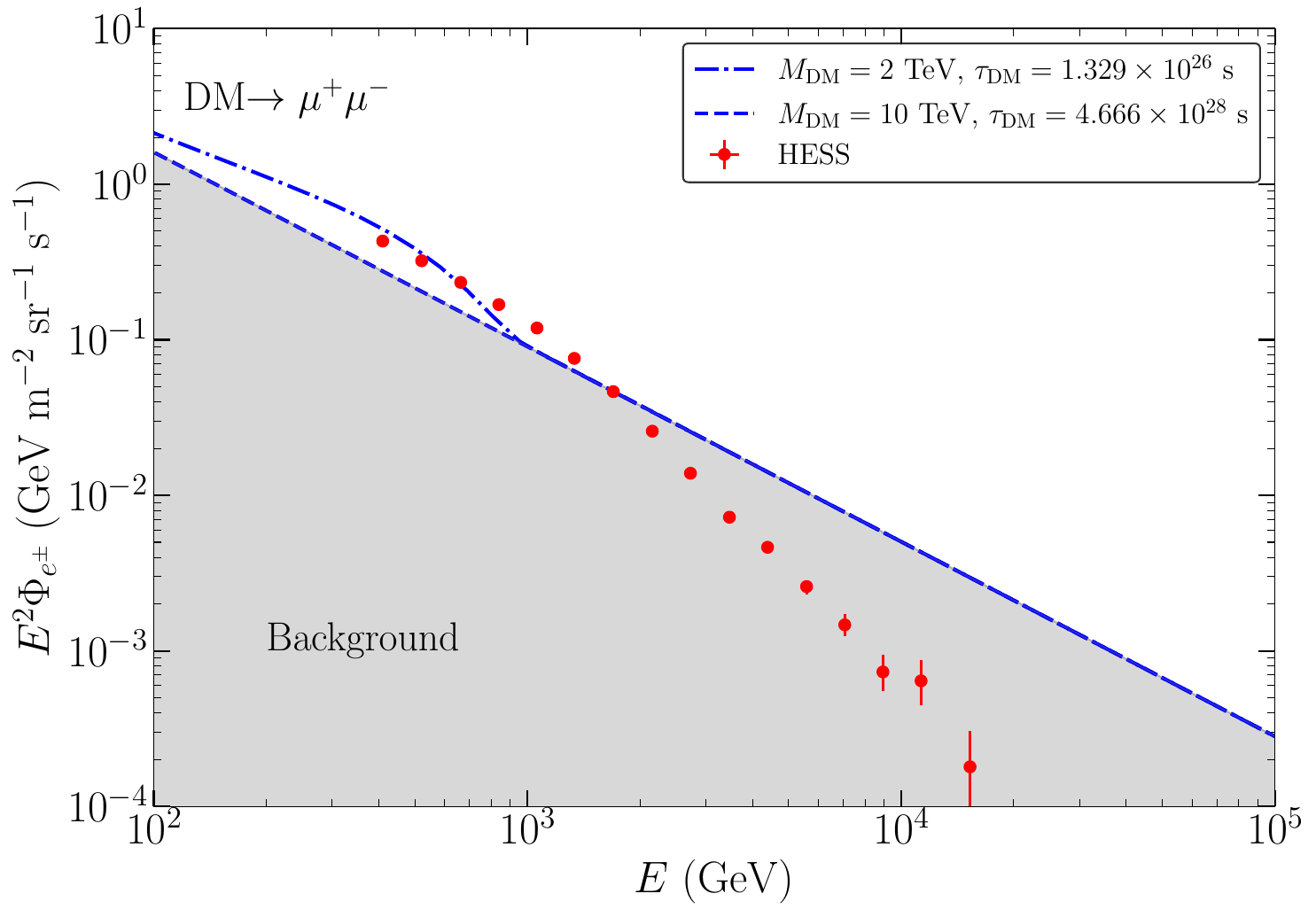}\caption{Same as in Fig. \ref{fig:caletflux} but for the HESS electron-positron flux.  The gray-shaded region is the electron-positron background given in Eq. (\ref{eq:bkg}).}
\label{fig:hessflux}
\end{figure}

In Fig. \ref{fig:hessflux}, we show the flux as a function of energy for two choices of DM masses for the HESS electron-positron flux measurements.

\subsection{HAWC}

\begin{figure}[H]
\centering
\includegraphics[scale=0.3]{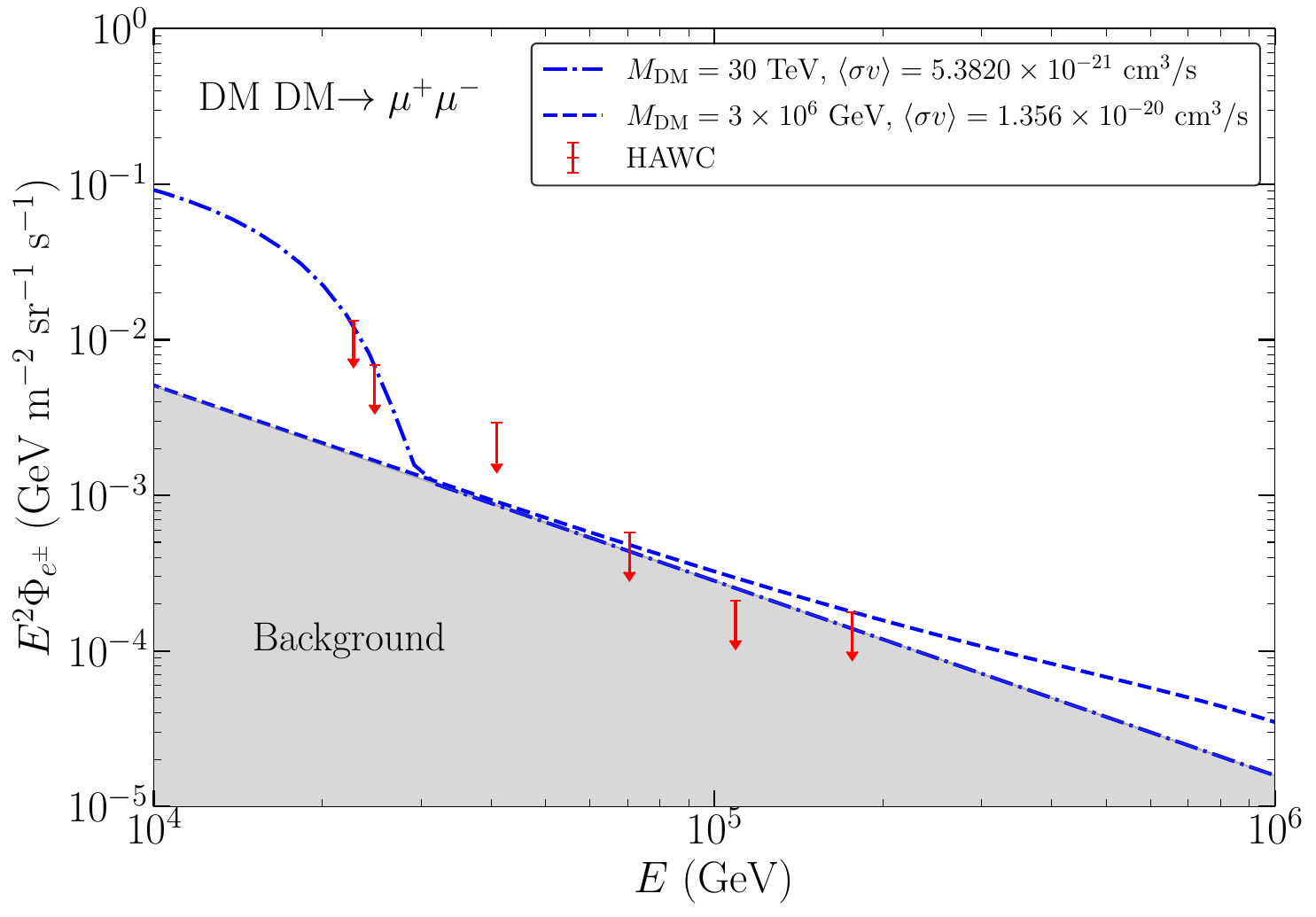}    \includegraphics[scale=0.3]{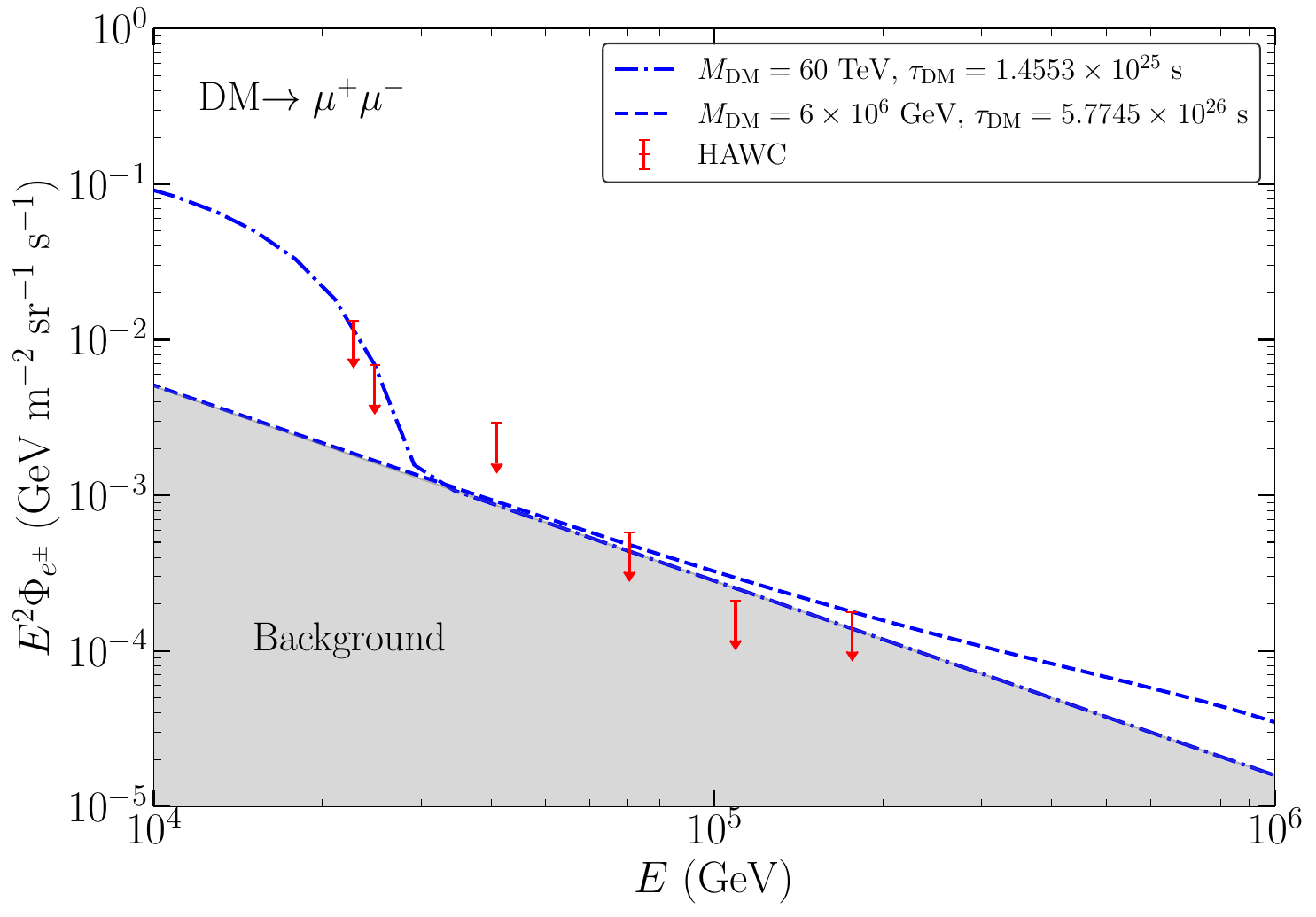}\caption{Same as in Fig. \ref{fig:hessflux} but for the HAWC $\gamma$-ray flux upper limits.}
\label{fig:hawcflux}
\end{figure}

In Fig. \ref{fig:hawcflux}, we show the flux as a function of energy for two choices of DM masses for the HAWC $\gamma$-ray flux upper limits.

\subsection{GRAPES-3}

\begin{figure}[H]
\centering
\includegraphics[scale=0.3]{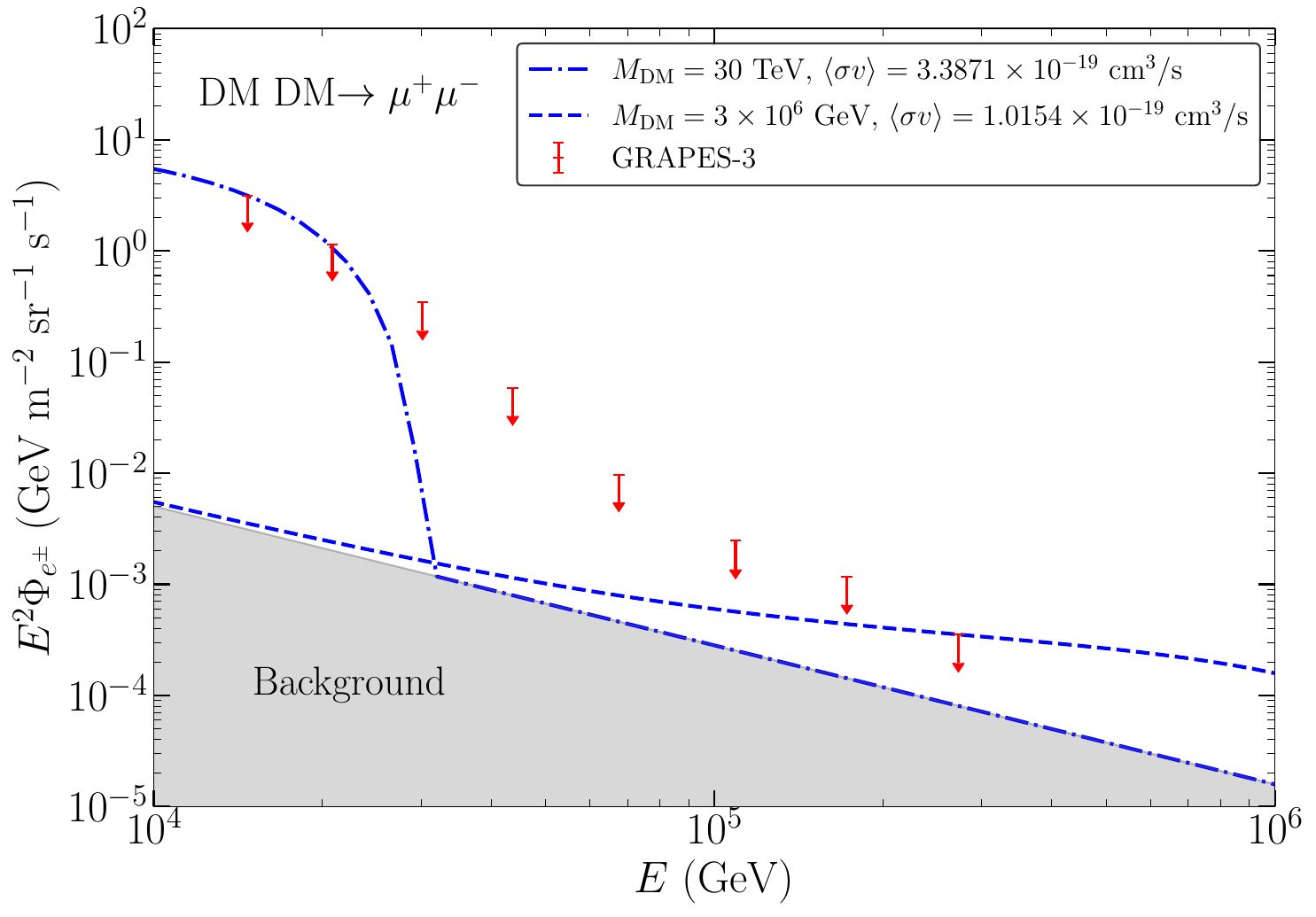}    \includegraphics[scale=0.3]{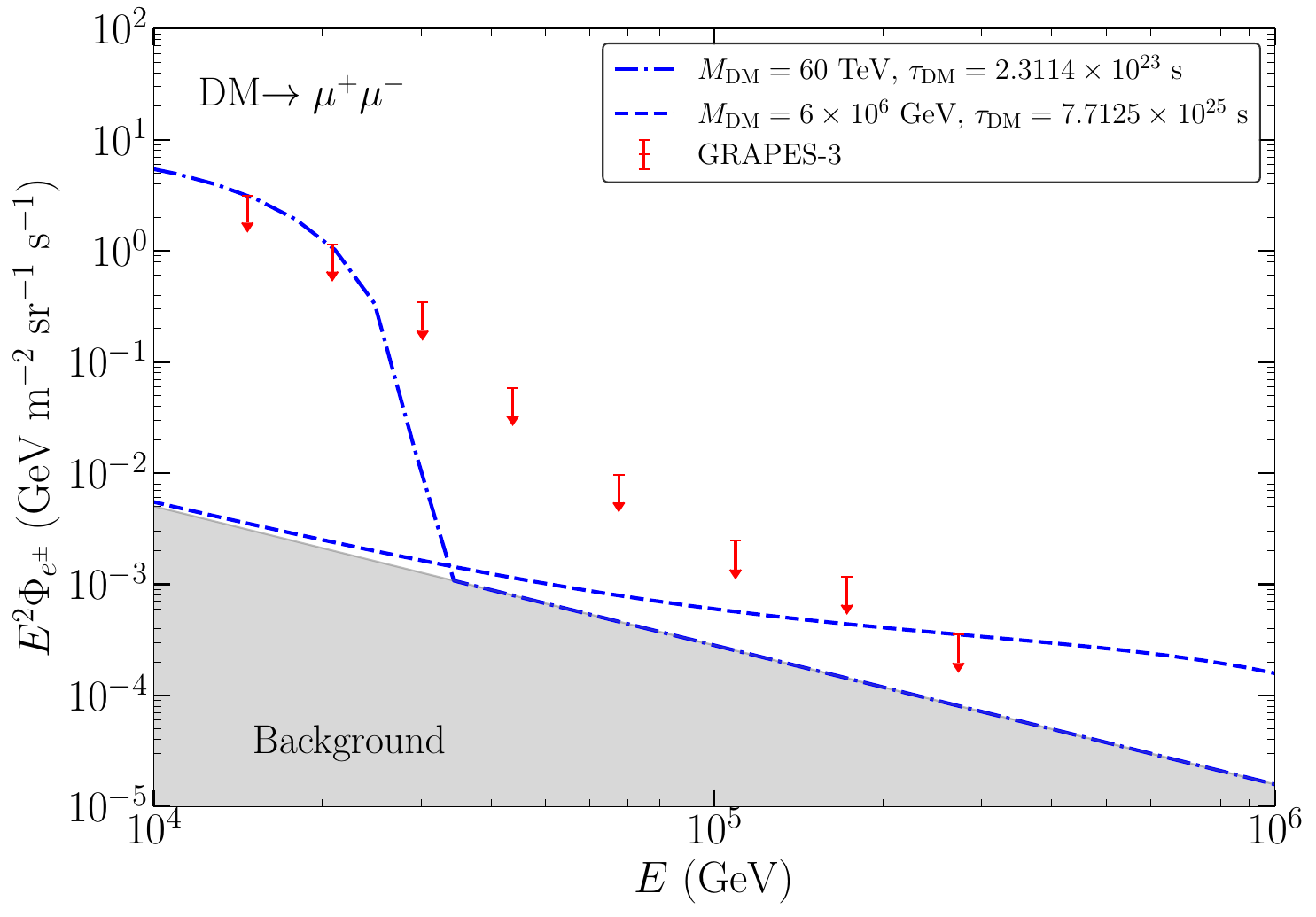}\caption{Same as in Fig. \ref{fig:hessflux} but for the GRAPES-3 $\gamma$-ray flux upper limits.}
\label{fig:grapesflux}
\end{figure}

In Fig. \ref{fig:grapesflux}, we show the flux as a function of energy for two choices of DM masses for the GRAPES-3 $\gamma$-ray flux upper limits.

\subsection{CASA-MIA}

\begin{figure}[H]
\centering
\includegraphics[scale=0.3]{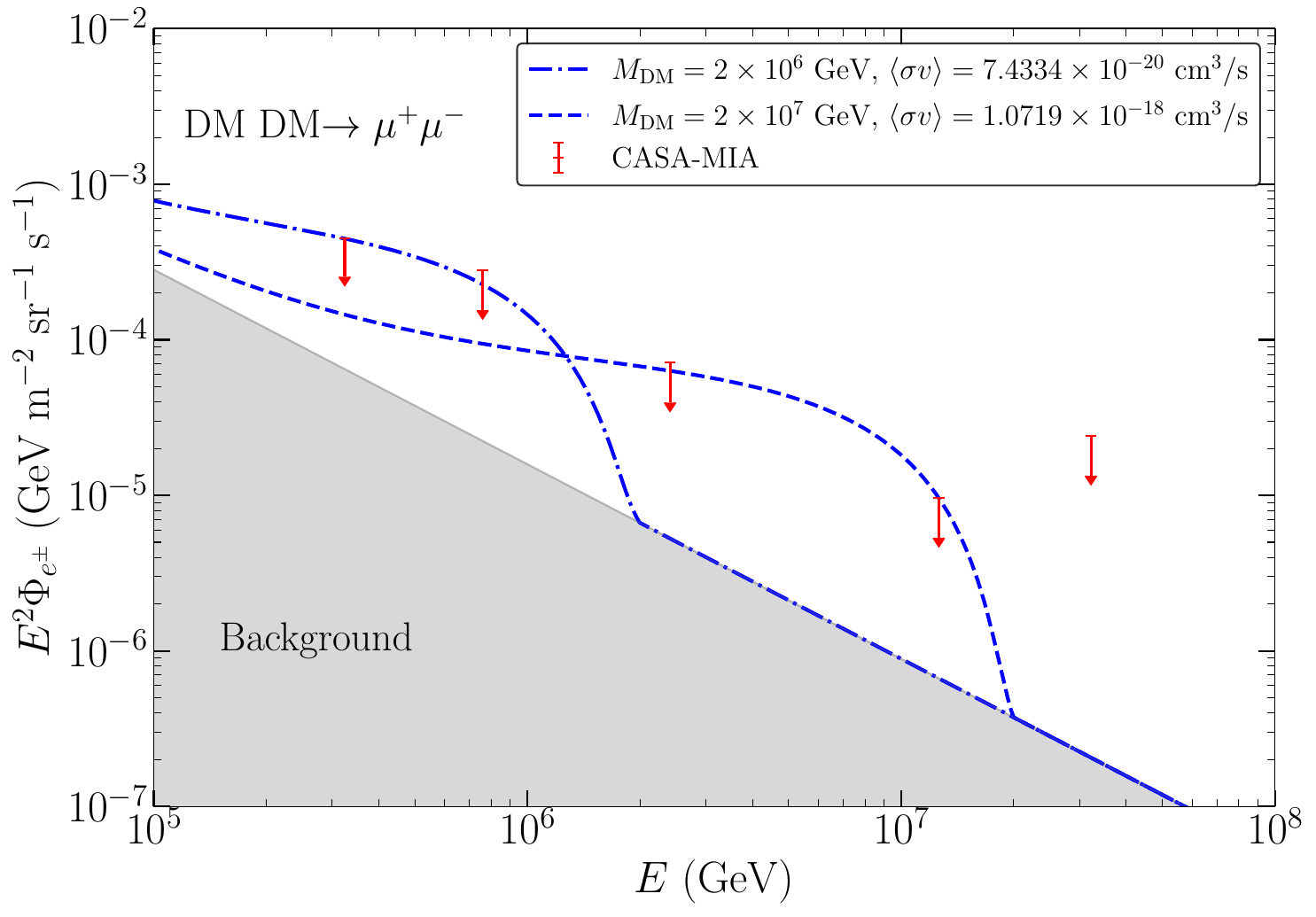}    \includegraphics[scale=0.3]{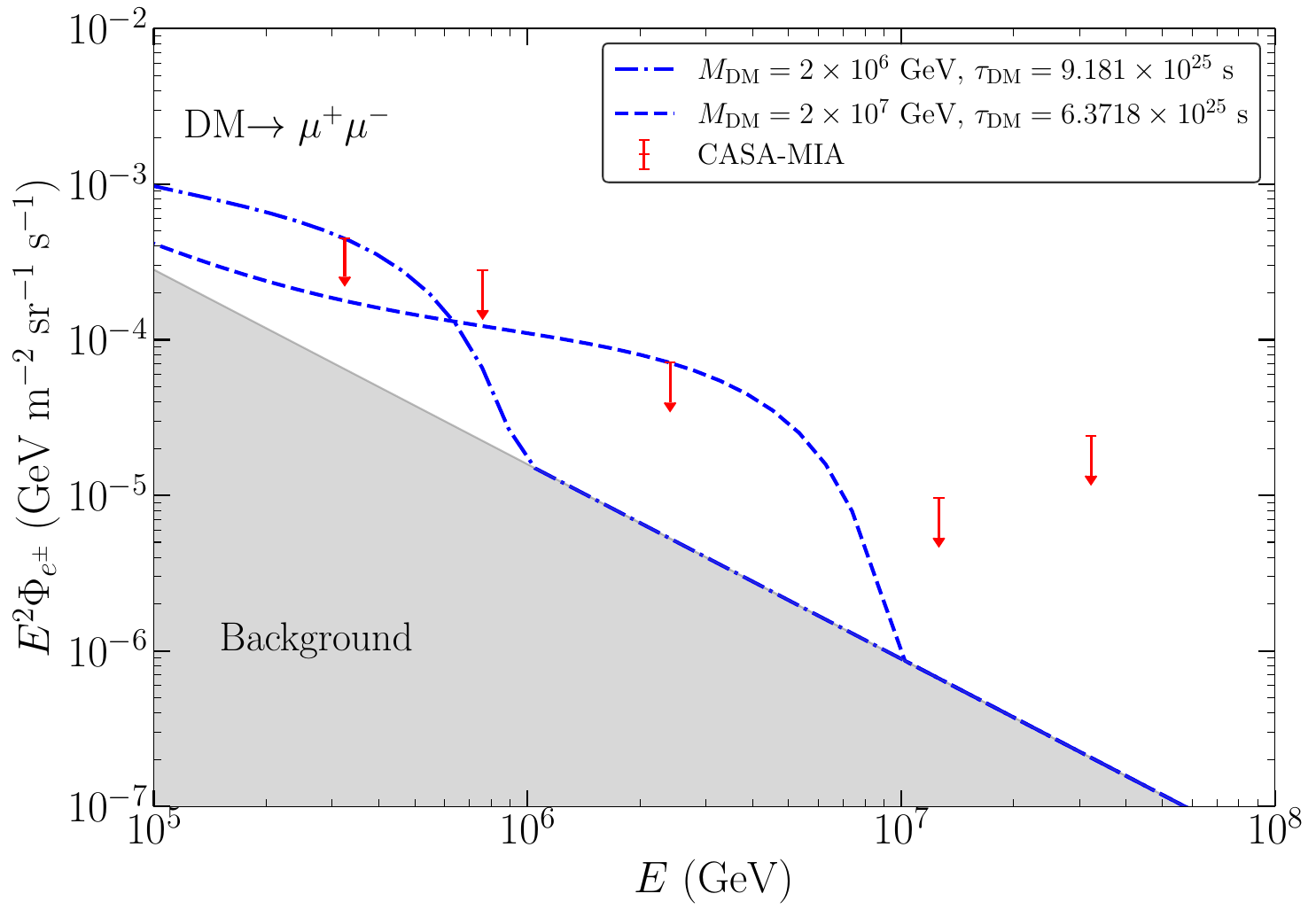}\caption{Same as in Fig. \ref{fig:hessflux} but for the CASA-MIA $\gamma$-ray flux upper limits.}
\label{fig:casamiaflux}
\end{figure}

In Fig. \ref{fig:casamiaflux}, we show the flux as a function of energy for two choices of DM masses for the CASA-MIA $\gamma$-ray flux upper limits.

\section{Annihilation cross-section and lifetime constraints for $b\bar{b}$, $e^+e^-$, $\mu^+\mu^-$, $\tau^+\tau^-$ channels}\label{app:otherchannel}
Here, we present the constraints on DM annihilation and lifetime for several other final states.
\begin{figure}[h]
	\centering	\includegraphics[scale=0.246]{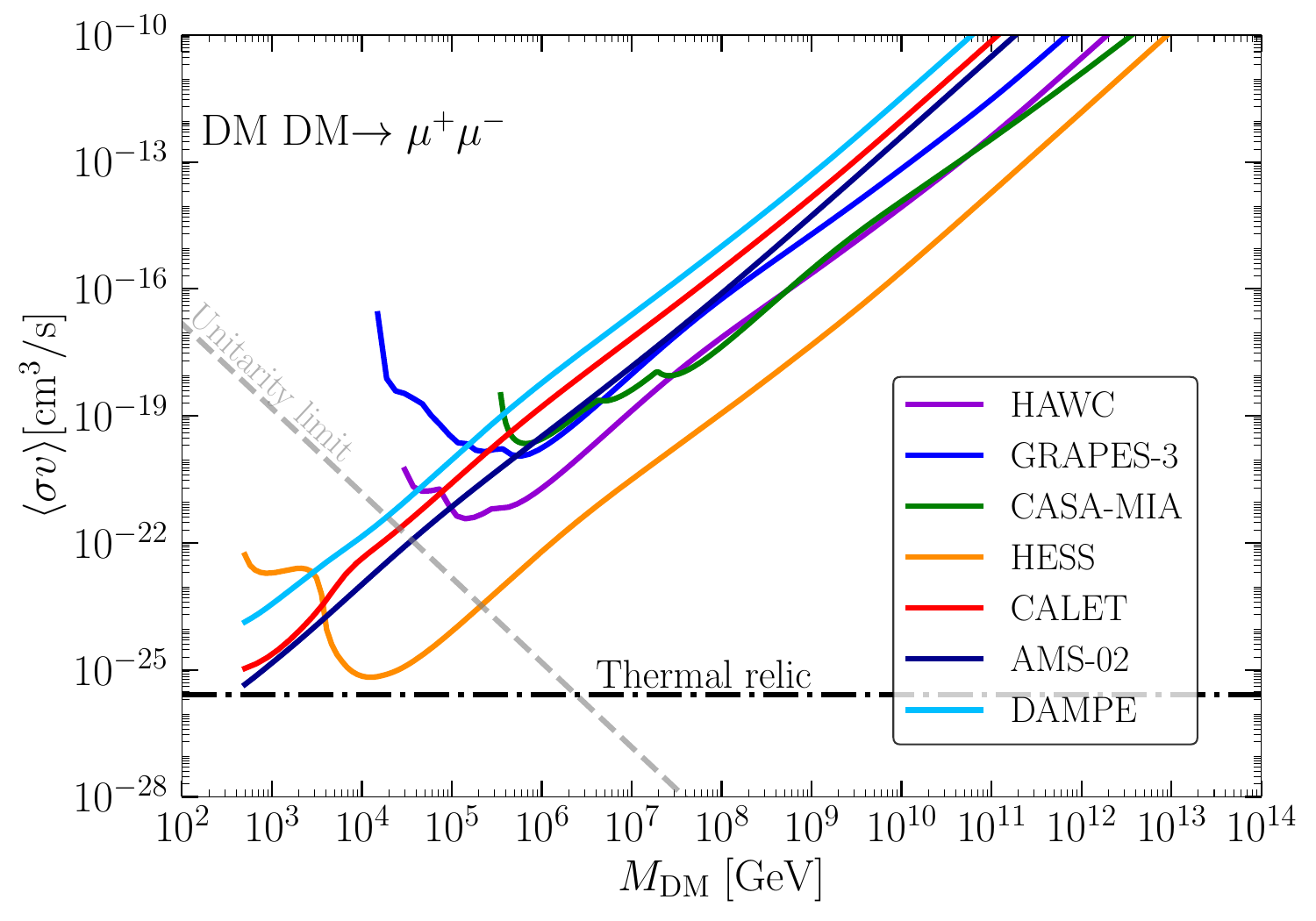}
   \includegraphics[scale=0.246]{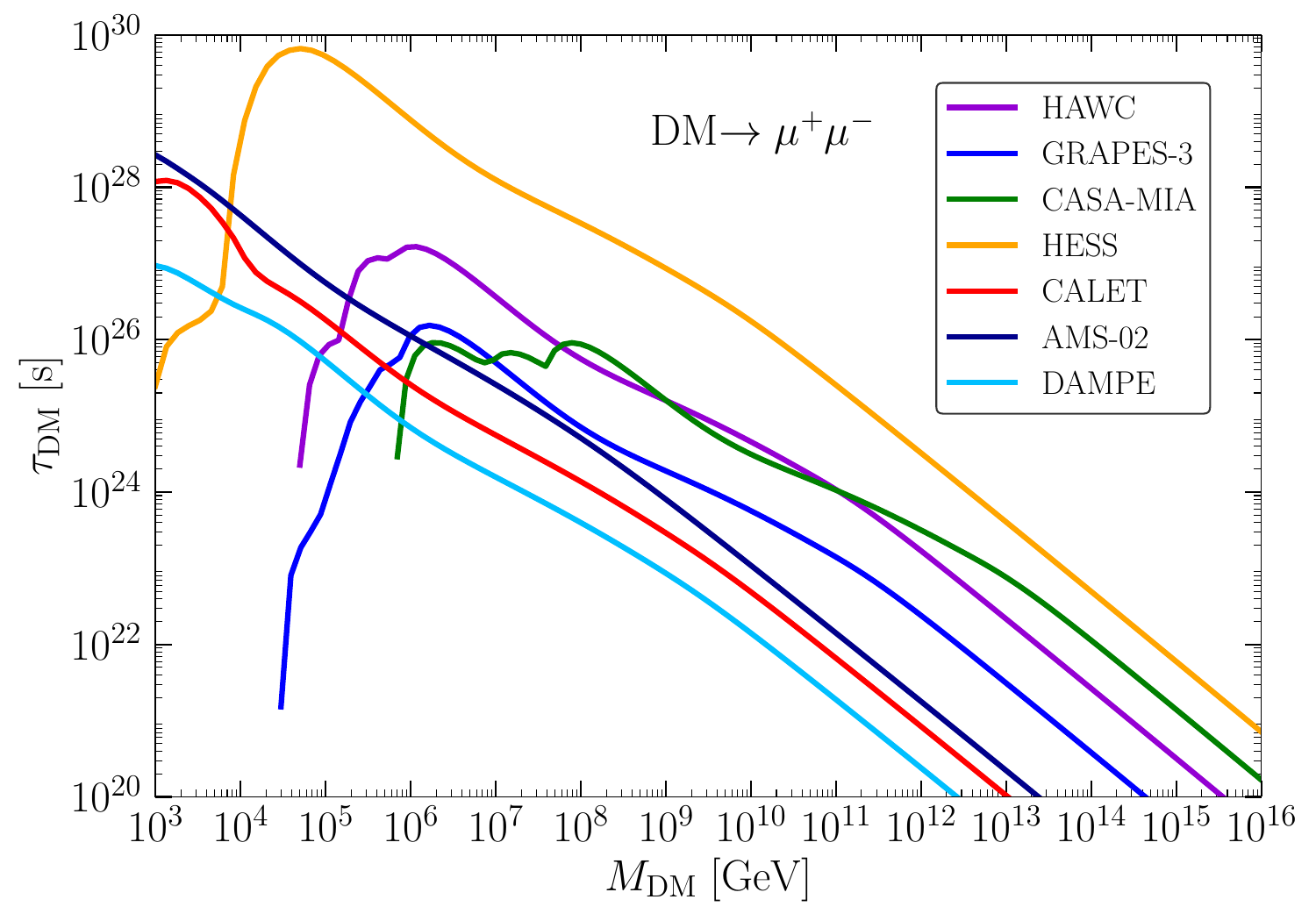}\\
    \includegraphics[scale=0.246]{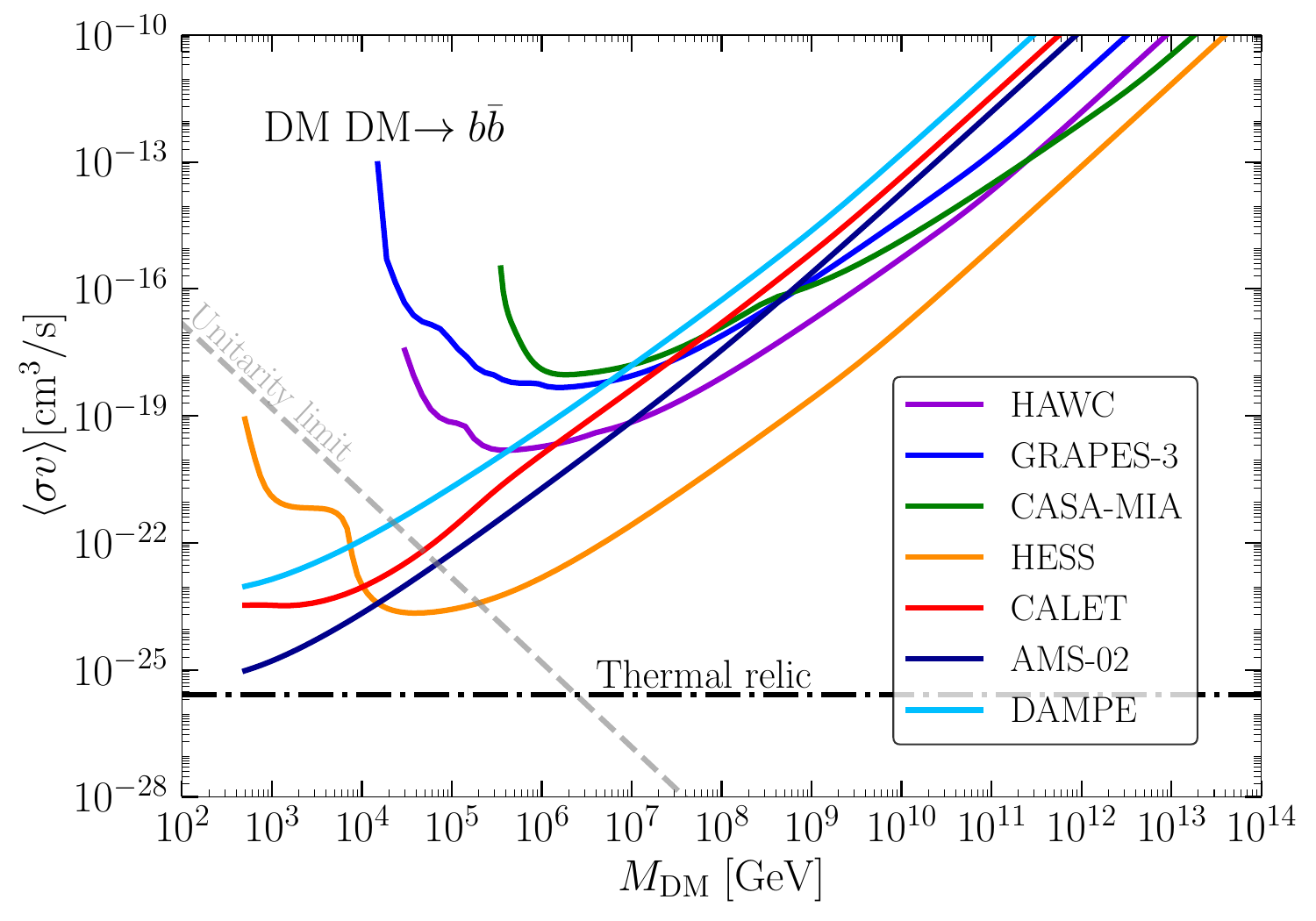}
    \includegraphics[scale=0.246]{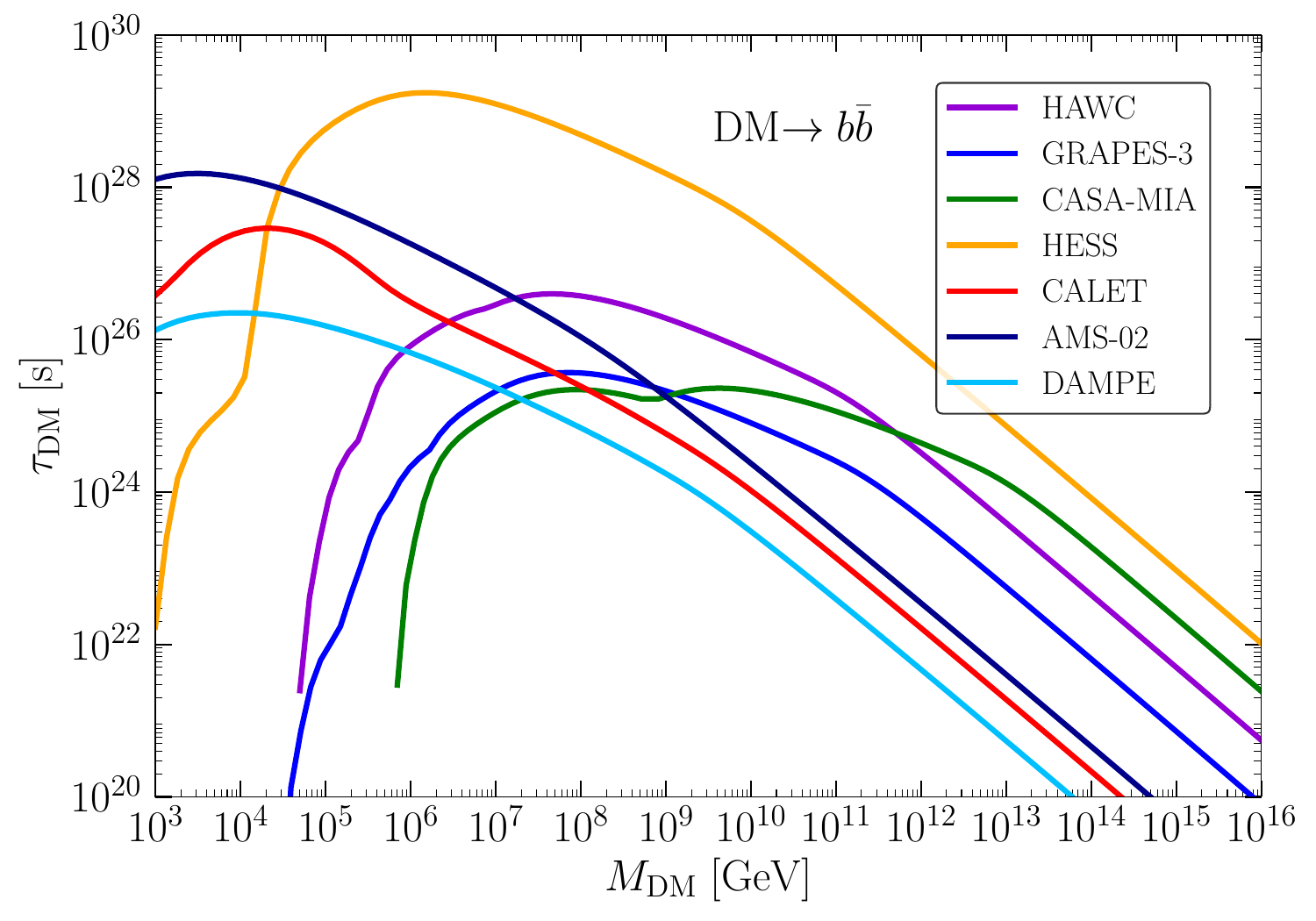}\\
     \includegraphics[scale=0.246]{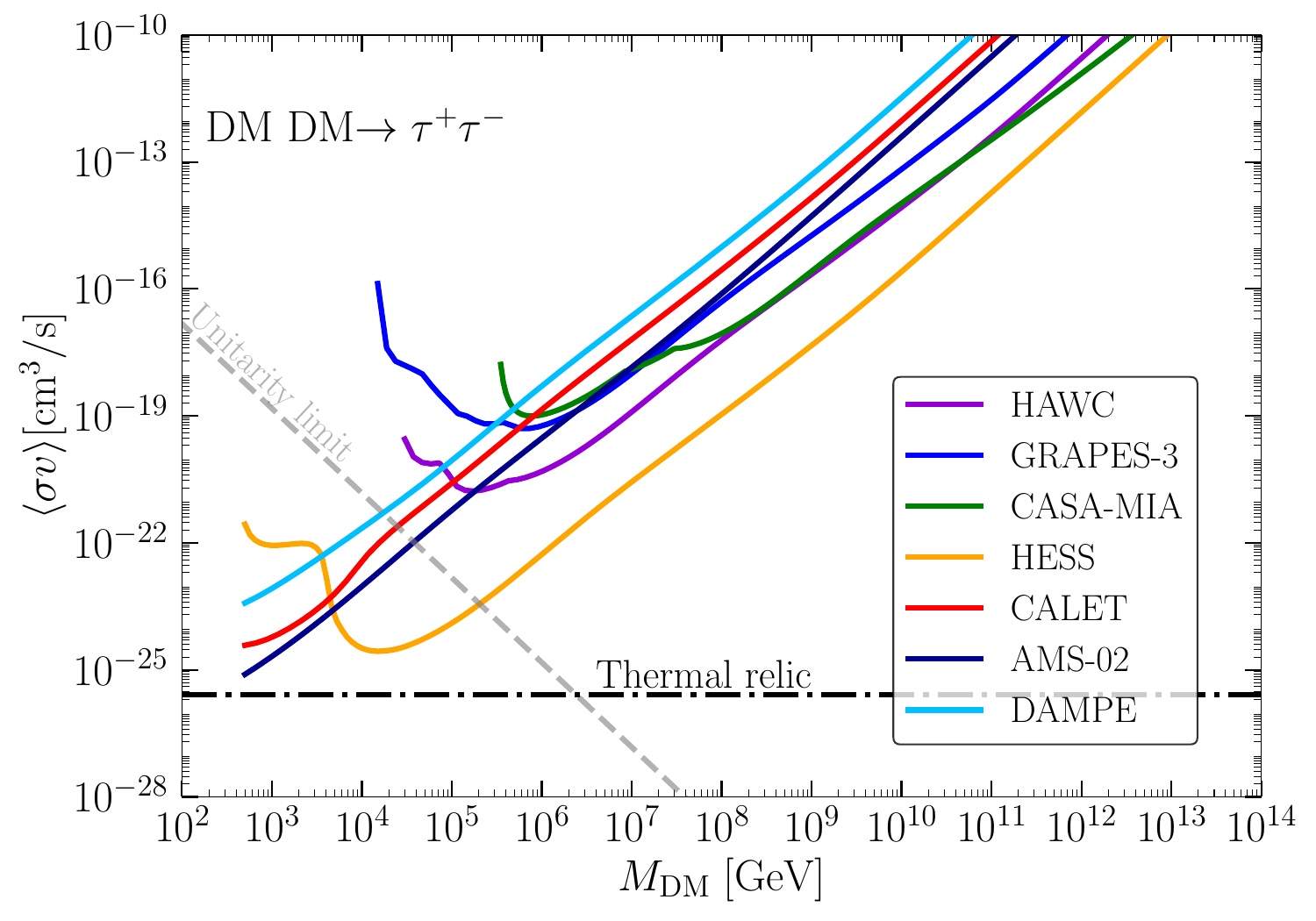}
    \includegraphics[scale=0.246]{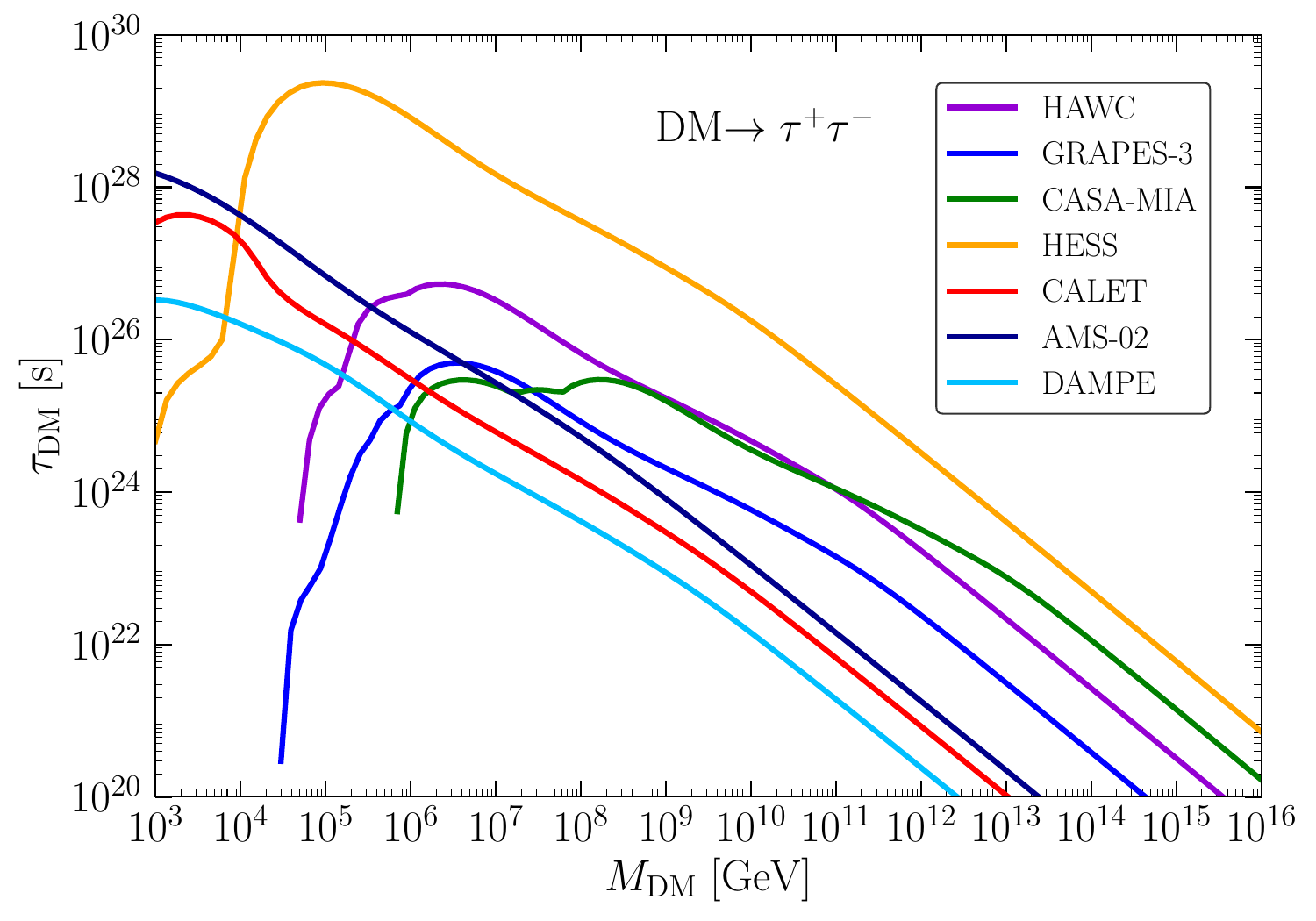}\\
    \includegraphics[scale=0.246]{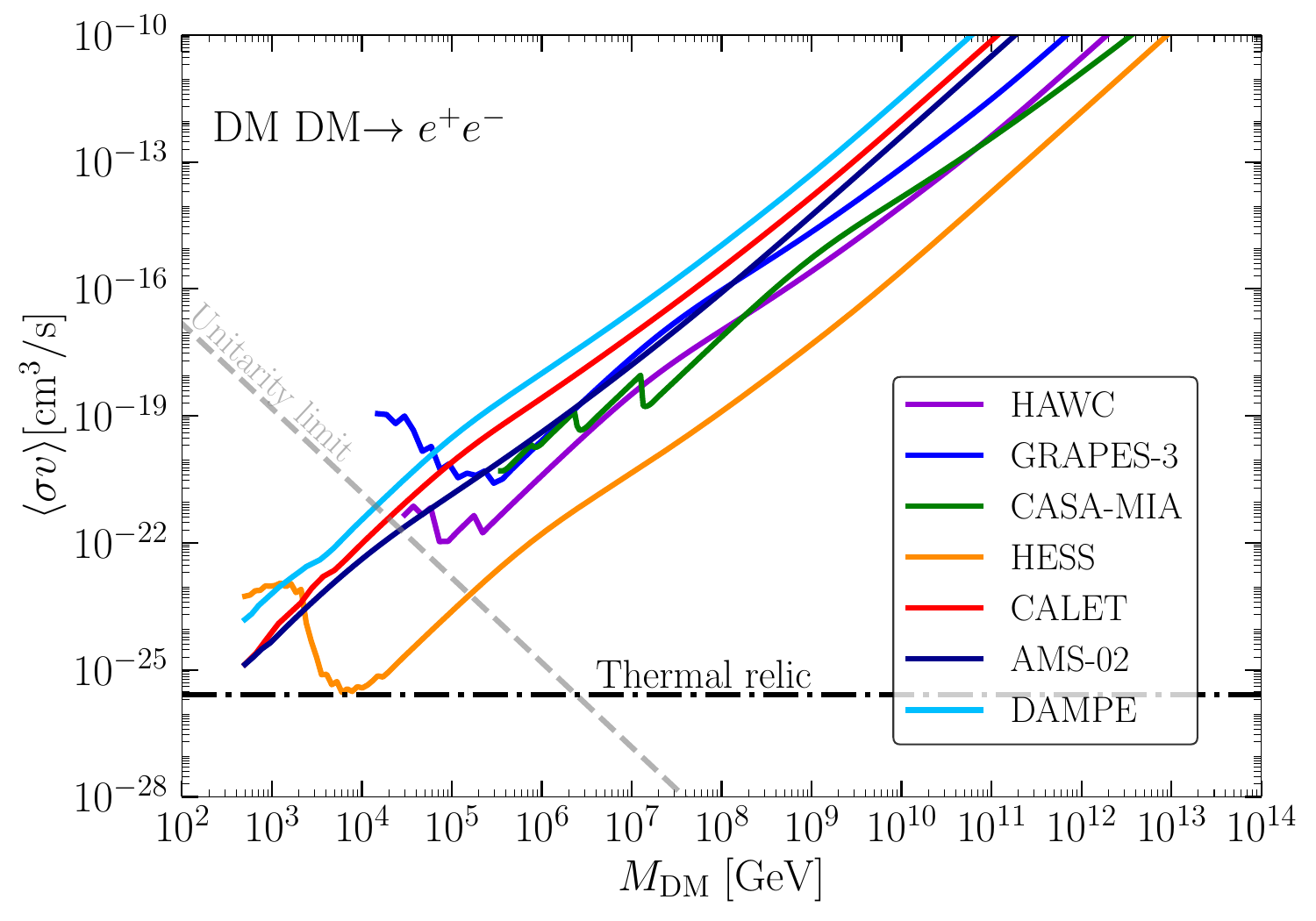}
    \includegraphics[scale=0.246]{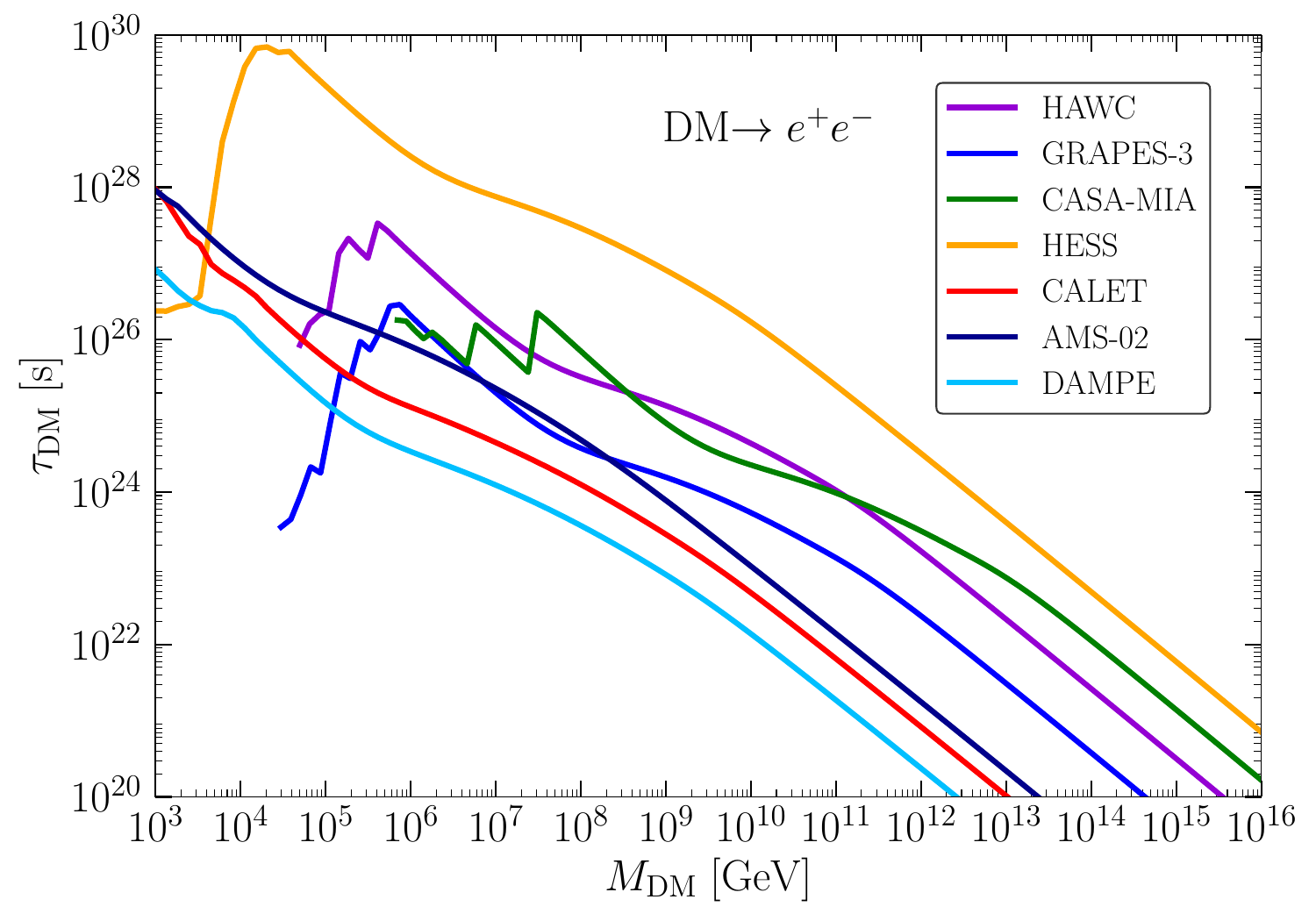}
	\caption{95\% C.L. limits on DM annihilation cross-section (\textit{left}) and DM lifetime (\textit{right}) from CALET, AMS-02, HAWC, GRAPES-3, CASA-MIA, DAMPE, and H.E.S.S. for various channels. From top to bottom, we show the cases of $\mu^+\mu^-$, $b\bar{b}$,  $\tau^+\tau^-$, and $e^+e^-$. Each line corresponds to a different experiment.}
	\label{fig:dmlimits}
\end{figure}


\providecommand{\href}[2]{#2}\begingroup\raggedright\endgroup

%
\end{document}